\documentclass[12pt]{article}

\usepackage{graphicx}
\usepackage{times}
\usepackage{amsmath,amssymb}
\usepackage{xcolor}
\usepackage[normalem]{ulem}
\usepackage{threeparttable}
\usepackage{notes2bib}
\usepackage[symbol]{footmisc}
\usepackage{hyperref}
\hypersetup{
    colorlinks=true,
    linkcolor=blue,
    filecolor=magenta,      
    urlcolor=cyan,
    pdftitle={Overleaf Example},
    pdfpagemode=FullScreen,
    }

\newenvironment{sciabstract}{%
\begin{quote} \bf}
{\end{quote}}

\title{Quantum state tracking and control of a single molecular ion in a thermal environment} 

\author
{
Yu Liu,$^{1,2,3\ast}$ Julian Schmidt,$^{1,2,4}$ Zhimin Liu,$^{1,2}$\\
David R. Leibrandt$^{1,2,5}$, Dietrich Leibfried$^{1,2}$, Chin-wen Chou$^{1,2}$\\
\\
\normalsize{$^{1}$Time and Frequency Division, National Institute of Standards and Technology,}\\
\normalsize{Boulder, Colorado 80305, USA}\\
\normalsize{$^{2}$Department of Physics, University of Colorado,}\\
\normalsize{Boulder, Colorado 80309, USA}\\
\normalsize{$^{3}$ Present address: Department of Chemistry and Biochemistry, University of Maryland,} \\
\normalsize{College Park, Maryland 20742, USA.} \\
\normalsize{$^{4}$ Present address: Paul Scherrer Institute (PSI), 5232 Villigen, Switzerland} \\
\normalsize{$^{5}$ Present address:  Department of Physics and Astronomy, University of California,}\\
\normalsize{Los Angeles, California 90095, USA} \\
\normalsize{$^\ast$ To whom correspondence should be addressed; E-mail: yuliu@umd.edu}
}

\date{}

\begin{document}


\baselineskip24pt


\maketitle



\noindent \textbf{One-sentence summary:} Real-time tracking of a single molecule’s quantum state enables the characterization and reversal of stochastic, environment-induced transitions.

\begin{sciabstract}

         Understanding molecular state evolution is central to many disciplines, including molecular dynamics, precision measurement, and molecule-based quantum technology. Details of the evolution are obscured when observing a statistical ensemble of molecules. Here, we reported real-time observations of thermal radiation-driven transitions between individual states (``jumps") of a single molecule. We reversed these ``jumps" through microwave-driven transitions, resulting in a twentyfold improvement in the time the molecule dwells in a chosen state. The measured transition rates showed anisotropy in the thermal environment, pointing to the possibility of using single molecules as in-situ probes for the strengths of ambient fields. Our approaches for state detection and manipulation could apply to a wide range of species, facilitating their uses in fields including quantum science, molecular physics, and ion-neutral chemistry.

\end{sciabstract}

The ability to follow the evolution of molecules at the level of their quantum states has revolutionized how we study the dynamics which occur in these systems. Resolving molecular states at progressively smaller energy scales often leads to the understanding of finer dynamical details. For example, resolving the vibrational states elucidate how energy flows within a molecule \cite{nesbitt1996vibrational} and in chemical reactions \cite{yang2007state}; resolving the rotational states provides insights into the interaction potential which governs molecular scatterings \cite{andres1980anisotropic, hutson1990intermolecular}; further resolving the spin-rotational states reveals the subtle role magnetic interactions play in intramolecular dynamics \cite{milner2014coherent} and reactions \cite{hu2021nuclear}. High resolution state detection have proven especially imperative for observing quantum mechanical effects in molecular dynamics such as resonance \cite{margulis2023tomography, kim2015spectroscopic}, interference \cite{chen2021quantum}, geometric phase \cite{yuan2018observation}, and entanglement \cite{li2019entanglement,liu2024quantum}. Continued progress in the experimental resolution of molecular states will provide new opportunities for the study, and ultimately the control, of molecular dynamics.

Over the past few decades, advances in atomic, molecular, and optical (AMO) physics enabled the manipulation and detection of pure molecular states in trapped, translationally cold neutral molecules. High-fidelity detection of individual molecular states are primarily accomplished via the fluorescence or absorption of photons, either on the molecules themselves \cite{shuman2010laser} or on atoms which are dissociated from these molecules \cite{ni2008high}. These capabilities have instigated efforts toward the broad application of molecules in quantum science and technology, including precision measurement \cite{acme2018improved}, quantum simulation \cite{christakis2023probing, li2023tunable}, and quantum computation \cite{holland2023demand, bao2023dipolar}. The use of these highly refined techniques to study molecular dynamics, however, remains challenging due to the need for repeated scattering of a large number of photons during detection (\textit{i.e.}, ``photon-cycling"), which requires a suitable level structure. Furthermore, the large translational energy deposition that occur in many molecular processes (\textit{e.g.}, reactions and collisions) cause Doppler broadening of transitions that render many states unresolvable.

Concurrent with the development of state manipulation and detection techniques for neutral molecules are those for molecular ions. In recent years, quantum-logic spectroscopy (QLS), a technique originally developed for atomic ion optical clocks \cite{schmidt2005spectroscopy}, has emerged as a new method for state preparation and measurement of single molecular ions \cite{wolf2016non,chou2017preparation,sinhal2020quantum}. 
With QLS, the quantum state information can be mapped between a molecular ion and a co-trapped, laser cooled, atomic ion via their coupled motion within an ion trap. In this way, QLS allows for projective preparation of the molecular ion in a pure quantum state as well as non-destructive state readout, all while leveraging the photon-cycling ability of the atomic ion. The atomic ion further provides efficient sympathetic cooling of the translational motions of the molecular ion to near their ground states, even in cases of large energy deposition into the translational motions \cite{hirzler2022observation}. Finally, both the translational cooling and QLS can be performed with relatively few requirements on the details of the molecular structure, and is therefore applicable to a wide variety of internal states and species. Numerous planned and ongoing experiments aim to leverage the precision and versatility of QLS for precision spectroscopy of a range of different species \cite{wellers2022controlled, zhou2024quantum, najafian2020megahertz, arrowsmith2023opportunities}. These features make QLS-controlled molecular ions a promising new platform to study molecular dynamics at the single particle level and with full state resolution.

In this study, we developed a QLS-based protocol to track and control the state evolution of a single CaH$^+$ molecular ion under environmental perturbations in a room temperature ultrahigh vacuum apparatus. Possible sources of perturbation include thermal radiation (TR) and background gas collisions. We observed transitions (``quantum jumps") between individual states of the molecule, and found them consistent with being TR-driven. By applying microwave pulses to drive rotational transitions in real time conditioned on the detected state, we reversed undesired state changes and confine the molecule within a target state for periods of $\sim20$ times its lifetime without such control.
The improved control over the state of the molecule increased the duty cycle with which operations such as spectroscopic transitions and quantum gates may be carried out from $\sim$7\% to $\sim$65\%. Measurements of transition rates between different molecular states suggested that the environment deviated from an ideal blackbody, demonstrating the potential of the molecule as a highly localized quantum sensor for its environment.

\section*{Quantum-logic detection of quantum jumps in CaH$^+$}

\begin{figure} [t!]
\centering
\includegraphics[width=1.00\textwidth]{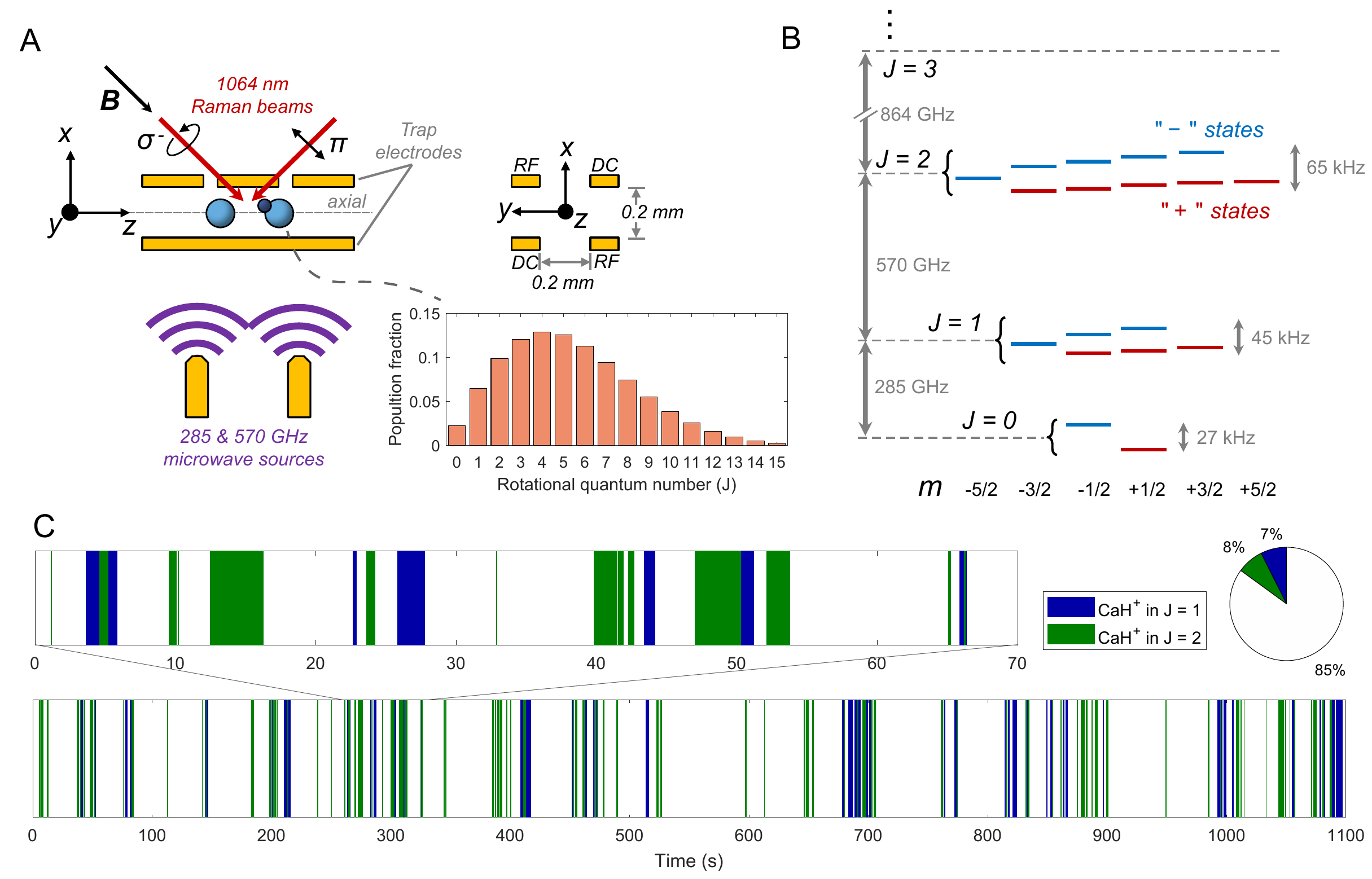}
    \caption{\small \textbf{Experimental setup and rotational dynamics in CaH$^+$} \textbf{(A)}  Schematic of the setup for quantum-logic (QLS) operation on CaH$^+$. A single $^{40}$Ca$^+$ and single $^{40}$CaH$^+$ are cotrapped in a linear RF trap and form a Coulomb crystal along the axial ($z$) direction. A magnetic quantization field of 6.5~G was directed diagonally in the $xz$ plane. Molecular transitions within each $J$-manifold were driven by a pair of 1064 nm laser beams in Raman configuration (red arrows), and those between different $J$-manifolds were driven by microwave radiations at hundreds of GHz (purple). A view of the trap along the axial direction shows the arrangement of the DC and RF electrodes. (Inset) The probability of finding the molecule in various rotational manifolds in the vibrational and electronic ground state of CaH$^+$, according to a 300 K Boltzmann distribution. \textbf{(B)} Energy level structure of CaH$^+$ for the first few rotational manifolds in the vibrational and electronic ground state (not to scale) in the presence of a 6.5~G magnetic field (\textit{B}) and a co-aligned 1300 V/m residual trap RF electric field (SM). Blue (red) lines represent states with $\xi = -$ ($+$). \textbf{(C)} Real-time observation of molecular state changes between $J = 1$ (blue), $J = 2$ (green), and $J$ other than 1 or 2 (white). (Upper left panel) A zoomed-in view over a particular time period to show details. (Upper right panel) A pie chart showing the fractional time the molecule is found in each manifold.}
\label{figIntro}
\end{figure}

Our experimental setup is schematically shown in Fig. \ref{figIntro}A and described in detail elsewhere \cite{chou2017preparation}. In brief, an ion crystal consisting of a $^{40}$Ca$^+$ and a $^{40}$CaH$^+$ is confined in a linear RF trap at room temperature and under ultrahigh vacuum ($\lesssim 10^{-8}$ Pa). Several coupled modes of the translational motion of the ions are cooled to the ground state via a combination of Doppler cooling, electromagnetically-induced transparency (EIT) cooling, and resolved sideband cooling. All cooling steps leverage the precise control over the internal states of Ca$^+$, and do not involve those of CaH$^+$. 
Internal degrees of freedom of CaH$^+$ may thermalize to the environment \cite{bertelsen2006rotational} through interaction with the thermal radiation (TR) emitted by surrounding surfaces, or collisions with residual background gas molecules (predominantly H$_2$). These interactions result in a distribution in the probability of finding the molecule in various rotational manifolds, labeled by the quantum number $J$ (Fig. \ref{figIntro}A inset). Within a given $J$-manifold, the probability is further divided evenly among $4J + 2$ spin-rotational sublevels $|\mathcal{J}\rangle \equiv |J, m, \xi\rangle$, where $m$ is the sum of the quantum numbers for the nuclear spin and rotational angular momentum projections ($m_I$ and $m_J$) along the quantization axis defined by a laboratory magnetic field (\textit{\textbf{B}} in Fig. \ref{figIntro}A), and $\xi \in \{ +, -\}$ indicates the relative sign in the superposition of product states with the same $m$ but opposite nuclear spin, \textit{i.e.}, $|J, m, \pm \rangle \equiv c^- |J, m_J = m+1/2 \rangle |m_I = -1/2\rangle \pm c^+ |J, m_J = m-1/2 \rangle |m_I = +1/2\rangle$, with $c^-$, $c^+>0$. 
For the extreme sublevels $|J, m = \pm(J + 1/2) , \pm \rangle = |J, m_J = \pm J \rangle | m_I 
 = \pm 1/2 \rangle$, which are simple product states, $\xi$ indicates the sign of $m$ \cite{chou2017preparation}.
 Fig. \ref{figIntro}B shows the spin-rotational level structure of CaH$^+$ in its vibrational and electronic ground state ($v = 0$, $X ^1\Sigma$) in the presence of external fields applied during experiments described in this work (see caption). One may also view the thermal distribution in a time-dependent picture, in which the molecular state evolves under the influences of external perturbations, causing the observed state to change sporadically (\textit{i.e.}, quantum jumps). State changes driven by TR follow dipole selection rules $\Delta J = \pm 1$ and $\Delta m = 0, \pm 1$, and those driven by collisions can take a wide range of $\Delta J$ and $\Delta m$ values.

We used QLS to non-destructively observe the molecular state and track its evolution under external influences. All QLS operations involved a pair of 1064 nm laser beams which drove, in a far-off-resonance Raman configuration, transitions between neighboring sublevels within a $J$ manifold. To detect whether the molecule was in $J$, we first concentrated the probability of finding the molecule distributed among the $4J+2$ sublevels into the extreme sublevel $|J, m = -J - 1/2, - \rangle$ via pumping, and then attempted a projection via the transition $|J, m = -J - 1/2, - \rangle \rightarrow |J, m = -J + 1/2, - \rangle$ (see SM and ref. \cite{chou2017preparation}). Following an initial successful projection, the molecule could be repeatedly re-projected in $J$ until a quantum jump occurred, and coherent operations such as spectroscopy \cite{chou2017preparation,chou2020frequency,collopy2023effects} or entanglement \cite{lin2020quantum} could be performed on a known, pure initial molecular state between re-projections. Fig. \ref{figIntro}C shows real-time observation of the molecule undergoing quantum jumps in and out of $J = 1$ and 2 in an experimental sequence where we made repeated and alternating detection attempts in these two manifolds. Each time the molecule was projected into $J = 1$ ($J = 2$), it spends, on average, 1.5(2) (0.7(1)) s before leaving (throughout the article we used one standard deviation (1SD) error as the measurement uncertainty).  The fractional time the molecule spent in each manifold defines the maximum duty cycle, $D$, over which we can perform coherent manipulations. Over the sequence in Fig. \ref{figIntro}C, we measured $D \sim $ 7~\% for $J = 1$ and 8~\% for $J = 2$. In this work we did not attempt detection in $J \geq 3$, which collectively contained $\sim 80\%$ of the probability in a 300 K Boltzmann distribution (Fig. \ref{figIntro}A inset), due to current limitations on the pumping efficiency for manifolds with large numbers of sublevels (SM).

\section*{Mechanism behind the observed jumps}

\begin{figure} [t!]
\centering
\includegraphics[width=1.00\textwidth]{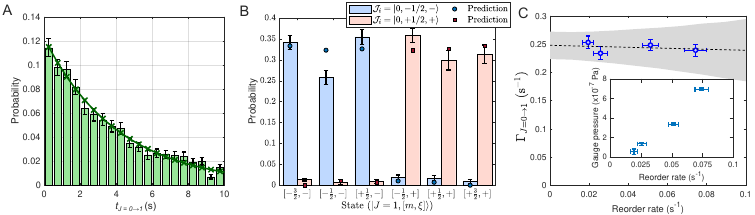}
    \caption{\small \textbf{Sublevel resolved quantum state jumps from J = 0 to 1} \textbf{(A)} Normalized histogram of durations between the molecule's initialization in $| 0, -1/2, - \rangle$ and its projection into any $| \mathcal{J} \in J = 1 \rangle$ sorted into 0.5 s time bins (light green bars). The data was obtained using the ``pump \& project" method. The histogram was fitted to an exponential decay, yielding a rate constant of $\Gamma_{J = 0 \rightarrow 1} = 0.244(8)$ s$^{-1}$. \textbf{(B)} Normalized histograms of observed jumps from either $| 0, -1/2, - \rangle$ (light blue bars) or $| 0, +1/2, + \rangle$ (light red bars) to each sublevel $| \mathcal{J} \in J = 1 \rangle$, and corresponding probabilities calculated assuming that the jumps conserved the nuclear spin of CaH$^+$ (dark blue circles and dark red squares). The data was obtained using the ``state-resolved" method. \textbf{(C)} Total rate for $J = 0 \rightarrow 1$ jumps, measured by ``pump \& project", for different reorder rates of the Ca$^+$-CaH$^+$ ion crystal. The data points (blue circles) were fitted to a linear function (black dashed line), yielding a slope of -0.09(0.25) and a vertical offset of 0.248(12) s$^{-1}$, consistent with no significant rate change due to background gas collisions in this pressure range. The light gray patch represents the 1 standard deviation (SD) confidence interval for the fit, obtained through parametric boostrapping. (Inset) Dependence of the ion reorder rate on the reading of a pressure gauge in the same vacuum system as the ion trap. All error bars presented in this figure represent 1 SD uncertainty.}
\label{figPopTrackJ0}
\end{figure}

To improve our control over the molecular state and increase the duty cycle of our experiments, we must first develop an understanding of the mechanisms driving the dynamics in our system. To this end, we monitored quantum jumps within the state space $J \in \{ 0, 1\}$, which is the minimal subspace within which effects of TR and collisions may be observed. 
We began each experiment by initializing CaH$^+$ in one of two sublevels in $J = 0$, $|\mathcal{J}_i \rangle = | 0, -1/2, - \rangle$ or $| 0, +1/2, + \rangle$. In anticipation of the molecule jumping to $J = 1$, we then performed detection in this manifold using one of two methods (SM): 1. sequentially attempting QLS projection from each of the 6 sublevels in $ J = 1 $ (referred to as the``state-resolved" method); 2. pumping the 6 sublevels towards and attempting QLS projection from the extreme state $|1, -3/2, - \rangle$ (referred to as the ``pump \& project" method). 
The attempts were repeated until a successful projection, after which the molecule was reinitialized to $|\mathcal{J}_i \rangle$ for another iteration. The results were collected over many iterations and summarized in Fig. \ref{figPopTrackJ0}.

Fig. \ref{figPopTrackJ0}A shows the normalized histogram of durations between the molecule's initialization in $| 0, -1/2, - \rangle$ and its projection into any $| \mathcal{J} \in J = 1 \rangle$, which we refer to as $t_{J = 0 \rightarrow 1}$. 
The distribution is well-described by an exponential function ($e^{-\Gamma_{J = 0 \rightarrow 1} t_{J = 0 \rightarrow 1}}$) with a fitted rate constant of $\Gamma_{J = 0 \rightarrow 1} = 0.244(8)$ s$^{-1}$, indicating that the observed jumps were a stochastic process with a mean rate of $\Gamma_{J = 0 \rightarrow 1}$.
Alternatively, we obtained $\Gamma_{J = 0 \rightarrow 1}$ by dividing the number of observed jumps by the total sequence duration over which the data in Fig. \ref{figPopTrackJ0}A was collected, and found its value to be $0.246(6)$ s$^{-1}$, consistent with the rate determined by the exponential fit. Fig. \ref{figPopTrackJ0}B displays the probabilities of detecting the molecule in $| \mathcal{J} \in J = 1 \rangle$ following its initialization in one of the two $| \mathcal{J} \in J = 0 \rangle $ states. Comparing the two sets of data, one observes that a molecule initially prepared in $| 0, -1/2 \rangle$ ($|0, +1/2 \rangle$) was predominantly found in the states with $\xi = - $ ($+$). At our operating magnetic field of 6.5 G, the spin and rotational angular momenta were reasonably decoupled, such that each sublevel had a dominant nuclear spin projection component (SM). In particular, the ``$-$'' (``$+$'') states have large amplitudes in the $m_I = -1/2$ ($+1/2$) component. As such, our observation suggested that the nuclear spin of the molecule was mostly preserved by the process causing the $J = 0 \rightarrow 1$ transitions. We compared the measured probabilities against those calculated assuming that the environment process coupled the rotational angular momenta ($|J, m_J\rangle$) but not the nuclear spins ($|I, m_I\rangle$) of the sublevels (SM), and found good overall agreements.

The observed nuclear spin conservation was consistent with the effects of TR, which drives electric dipole transitions. On the other hand, most inelastic collisions tend to leave nuclear spins unchanged as well. To investigate the relative contributions of these two mechanisms to the observed jumps, we measured the rate of jumps at different background pressures. For each pressure value, we determined the rate at which the Ca$^+$ and CaH$^+$ exchanged their positions in the crystal due to collisions, and used this ``reorder'' rate as an \textit{in situ}, relative measure of the total collision rate between the CaH$^+$ and background gas molecules \cite{hankin2019systematic}. The reorder rate was well-correlated to the reading of a nearby pressure gauge (inset of Fig. \ref{figPopTrackJ0}C). The results (Fig. \ref{figPopTrackJ0}C) showed no significant change in $\Gamma_{J = 0 \rightarrow 1}$ over a factor of $\sim 4$ variation in the reorder rate. A linear fit to the data bounded the contribution of collisions to $\Gamma_{J = 0 \rightarrow 1}$ to below 0.02 s$^{-1}$ at our nominal operating pressure of $\lesssim 10^{-8}$ Pa. 
We did not experimentally investigate the effect of pressure on the rates of quantum jumps between other pairs of rotational manifolds.
However, it is generally observed for diatomic molecules that probability for state-changing collisions tends to decrease for increasing initial rotational quantum number ($J_i$) or increasing difference to the final rotational quantum number ($\Delta J = J_f - J_i$) \cite{brunner1981rotational,kliner1999measurements}.
As such, collisions with background gas molecules are, in general, unlikely to be a significant cause of $J$-changing transitions in CaH$^+$ over the range of conditions explored here.
All subsequent experiments described here were carried out at or below a reorder rate of 0.02 s$^{-1}$.

\section*{Molecular state tracking and control}

\begin{figure} [pt!]
\centering
\includegraphics[width=1.00\textwidth]{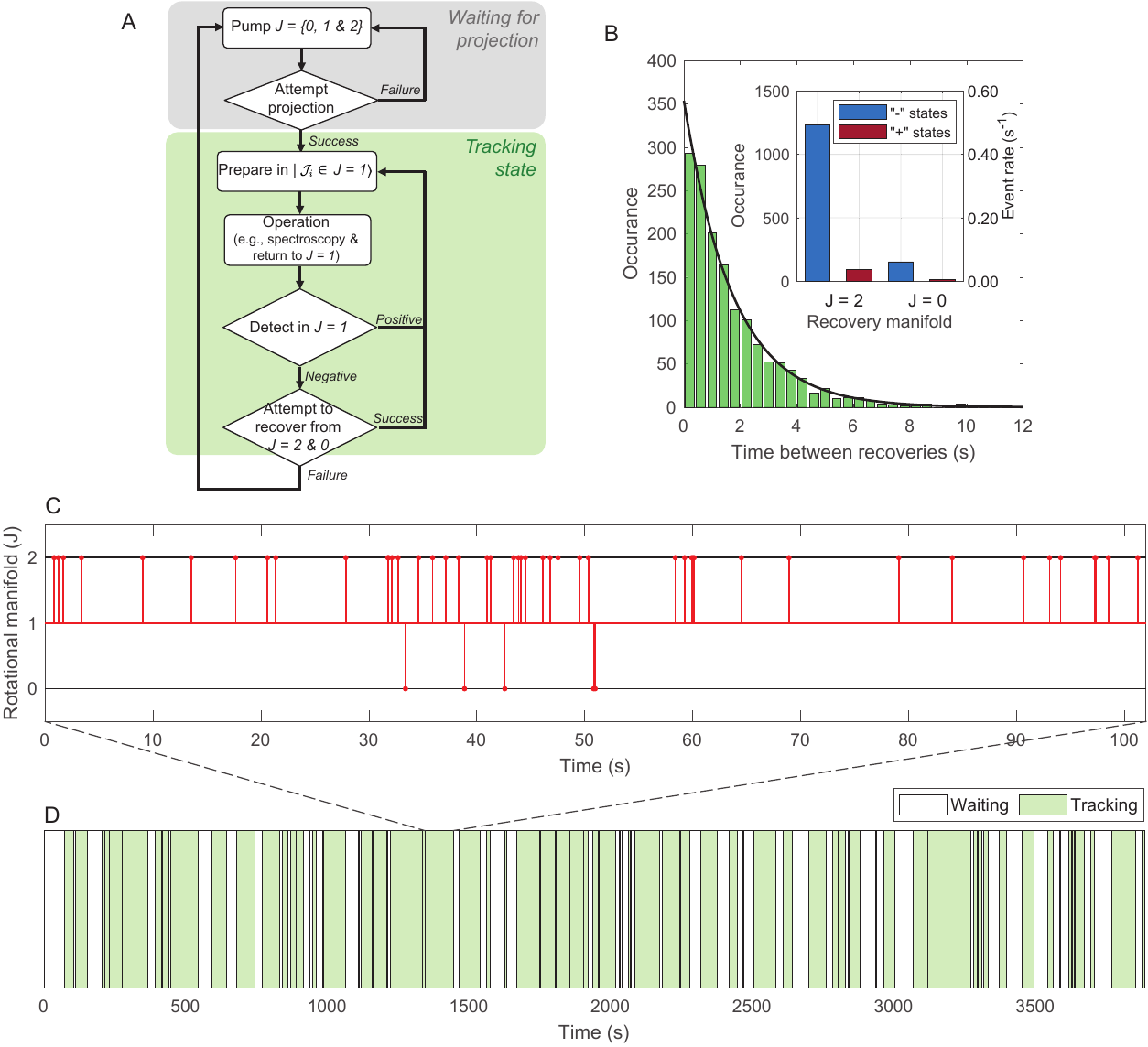}
    \caption{\small \textbf{Tracking and controlling the molecular state} \textbf{(A)} Flowchart for molecular state control. \textbf{(B)} Distribution of the time the molecule spent in $J = 1$ between two successive recoveries from $J = 0$ or 2 (green bars). A fit of the distribution to an exponential decay (black curve) yielded a $1/e$ time constant of $\tau_{J = 1} = 1.71(6)$ s. (Inset) Number (left y-axis) and rate (right y-axis) of recovery events associated with the ``$+$'' and ``$-$'' states of $J = 0$ and 2. Because the molecule was prepared in $|\mathcal{J}_i \rangle = | 1, -1/2, -\rangle$, it was driven by TR predominantly to other ``$-$'' states. \textbf{(C)} Real-time observation of quantum jumps from $J = 1$ to $J = 0$ and 2. The molecular state at any point in time is indicated in red. Each vertical line marks a recovery event, which contained both the jump out of and recovery back to $J = 1$. Because the average time for a recovery was on order of 10 ms (SM), the two processes were not resolvable on the timescale of this plot. \textbf{(D)} Evolution of the molecule between the tracked ($J \in \{0,1,2\}$) and untracked ($J > 2$) subspaces during the execution of the protocol, marked by light green and white patches, respectively.}
\label{figPopTrackJ1}
\end{figure}

Having established TR as predominantly responsible for the observed quantum jumps between different $J$-manifolds, we designed a protocol to undo these changes and keep the molecular state within a target manifold $J_i$. 
Leveraging the $\Delta J = \pm 1$ selection rule which TR-driven transitions obey, we tracked quantum jumps from $J_i = 1$ to $J = 0$ and 2, and reversed them using electric dipole transitions driven by microwave $\pi$-pulses.
The experiment began with repeated detection attempts in $J \in \{0, 1, 2\}$, during which we waited for TR to drive the molecule into this state space. Following a positive projection, we transferred the molecule to a certain sublevel $| \mathcal{J}_i \in J = 1 \rangle $, and began performing the desired operation which generally entailed some coherent manipulation of the molecular state. In case the operation took the molecule out of $J = 1$ (\textit{e.g.}, spectroscopic probe of an excited rovibrational level), we returned the state to $J = 1$ afterwards. 
We then detected whether the molecule still resided in $J = 1$, and, if so, transferred the state back to $| \mathcal{J}_i\rangle $ for another iteration of the operation. Failure to detect in $J = 1$ triggered a recovery attempt that searched the neighboring manifolds $J = 0$ and 2.
The search began in $J = 2$, which represented an effective ``border" of the state space we were working in, above which our state detection became less efficient (SM). 
If the molecule was detected in $J = 2$, we drove it back to $J = 1$ using a $\sim$570 GHz microwave $\pi$-pulse; otherwise we detected the two sublevels of $J = 0$ by coherently transferring any state amplitude in $|0, -1/2, - \rangle$ ($|0, 1/2, + \rangle$) to $| 1, -3/2, - \rangle$ ($| 1, -1/2, - \rangle$) via a $\sim$285 GHz microwave $\pi$-pulse and then attempting a projection. A direct QLS detection in $J = 0$ is currently not feasible because its two sublevels are not coupled by the 1064 nm Raman beams used for QLS operations (SM). If the molecule was successfully recovered back to $J = 1$, the experiment resumed; otherwise, tracking of the state was unsuccessful and we must wait for the molecule to re-enter $J \in \{ 0, 1, 2\}$ after a period of uncontrolled evolution. A flowchart for the protocol is provided in Fig. \ref{figPopTrackJ1}A, and a timing diagram is provided in Figure S3.

To evaluate the effectiveness of this state control protocol, we executed it continuously for a duration of approximately one hour. For the purpose of this evaluation, the operation on the molecule was simply a 25 ms wait and the molecule was in the state $|\mathcal{J}_i \rangle = | 1, -1/2, -\rangle$. The results are summarized in Fig. \ref{figPopTrackJ1}.
Fig. \ref{figPopTrackJ1}C shows TR-induced quantum jumps from $J = 1$ to $J = 0$ and 2 observed during a particular tracking period, which begans with an initial preparation in $J = 1$ and ended with the failure to recover from $J = 0$ and 2. 
Fig. \ref{figPopTrackJ1}B shows a histogram for the duration the molecule spends in $J = 1$ before jumping to a neighboring manifold. Fitting the histogram to an exponential decay, we found the lifetime of $J = 1$ to be $\tau_{J = 1} = $ 1.71(6) s. 
The inset displays the number of times the molecule was recovered from $J = 0$ and 2 (left y-axis). The results were separately tallied for the ``$+$'' and ``$-$'' states of each manifold.
Dividing the number of recovery events by the total time the molecule spent in $J = 1$, we obtained the rates of the TR-induced transition from $J = 1$ to the neighboring manifolds (right y-aixs). 
Over the execution of the protocol, the system alternated between waiting for an initial projection and tracking the state (Fig. \ref{figPopTrackJ1}D). 
The average waiting period was $\overline{T}_{\text{wait}} = 19.3$ s, and the average tracking period was $\overline{T}_{\text{track}} = 35.5$ s. Thus we found the duty-cycle, defined as the fraction of time over which we could confine the molecule to $J = 1$, to be $D = \overline{T}_{\text{track}}/(\overline{T}_{\text{track}} + \overline{T}_{\text{wait}}) = 64.7\%$. This result represented an improvement of about one order of magnitude compared to when the state was not actively controlled (Fig. \ref{figIntro}C). Mechanisms which limited the performance of our protocol are discussed in the SM.

\section*{Characterizing the thermal radiation environment with the molecule}

\begin{figure} [t!]
\centering
\includegraphics[width=0.80\textwidth]{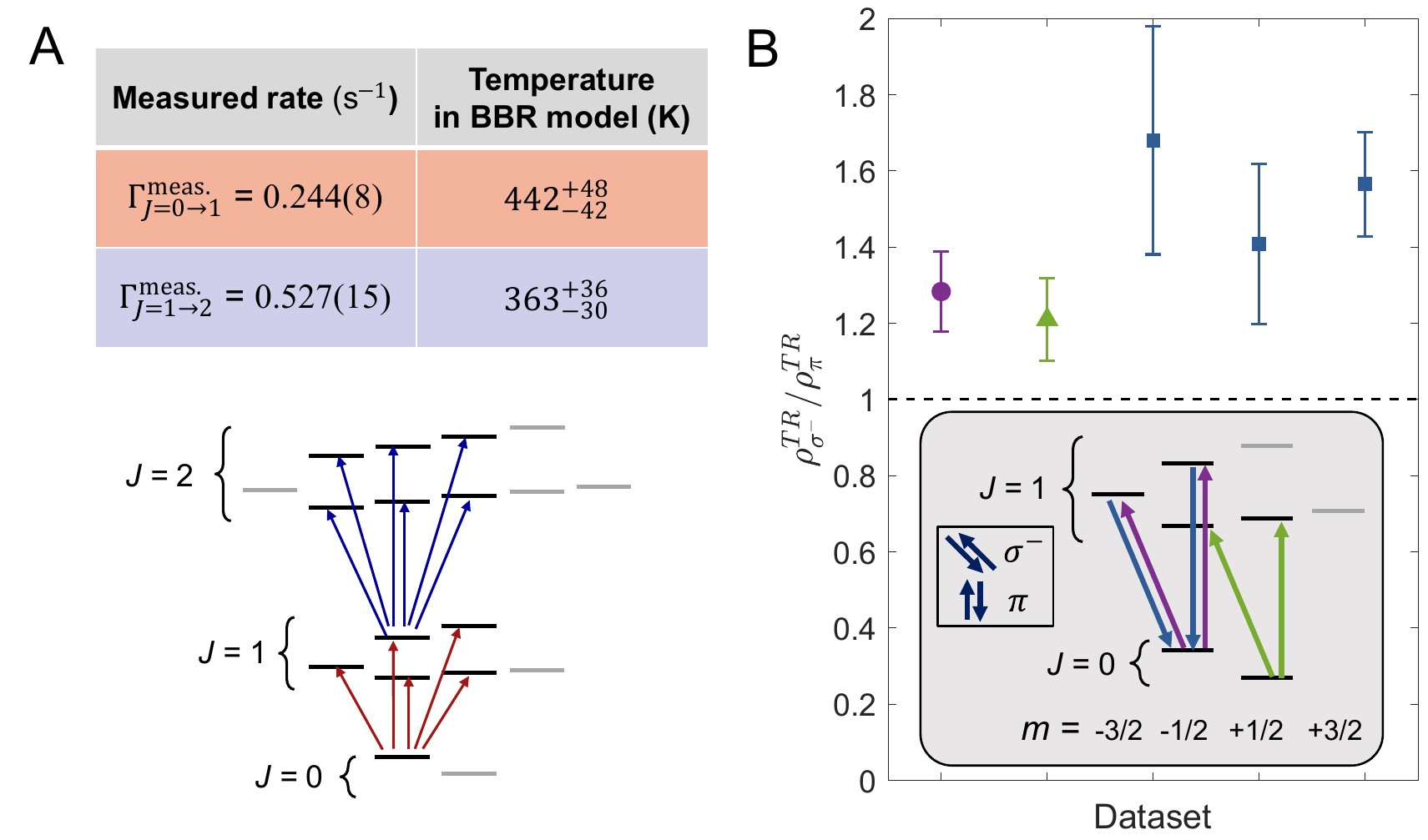}
    \caption{\small \textbf{Deviations from a blackbody environment}
    \textbf{(A)} (Upper panel) Measured rates for $J = 0 \rightarrow 1$ and $J = 1 \rightarrow 2$ transitions and the corresponding temperatures of a blackbody environment ($T$). Uncertainties in $T$ were derived from the uncertainties of both $\Gamma^{\textrm{meas.}}$ and the permanent dipole moment of CaH$^+$ (5.34 $\pm$ 0.19 Debye \cite{collopy2023effects}). Because the measured rates underestimated the actual transition rates, the values for $T$ should be considered as lower bounds. (Lower panel) Individual transitions which contributed to $\Gamma^{\textrm{meas.}}_{J = 0 \rightarrow 1}$ ($\Gamma^{\textrm{meas.}}_{J = 1 \rightarrow 2}$), marked by red (blue) arrows. \textbf{(B)} Ratios between the energy densities of $\sigma^{-}$- and $\pi$-polarization components of TR ($\rho^{\textrm{TR}}_{\sigma^{-}}$ and $\rho^{\textrm{TR}}_{\pi}$) at around 285 GHz, derived from measurements on the three pairs of transitions indicated in the inset. The purple circle and green triangle points derive from the $J = 0 \rightarrow 1$ transition probabilities presented in Fig. \ref{figPopTrackJ0}B. The blue square points derived from the rates of $J = 1 \rightarrow 0$ transitions obtained during the tracking experiment summarized in Fig. \ref{figPopTrackJ1}, with the 25 ms wait applied and the molecule was in $| 1 ,-3/2, - \rangle$ or $| 1 ,-1/2, - \rangle$. The rate data for the blue square points were collected over three sets on different days. Error bars represent 1 SD uncertainty.}
\label{figBBRCompare}
\end{figure}

For the discussions that follow, we emphasize the distinction between the terms ``thermal radiation" (TR) and ``blackbody radiation'' (BBR). Through out this article, TR refers to the electromagnetic radiation emitted due to the stochastic motion of particles in materials. TR which are emitted by an ideal blackbody and therefore follow Planck's law are referred to specifically as ``blackbody radiation'' (BBR). The spectrum of TR from realistic sources can differ from that of BBR due to scattering or boundary conditions.

In many experiments, including trapped ion optical atomic clocks \cite{huntemann2016single,brewer2019al+}, the interaction between particles and TR is a concern. During the tracking experiments summarized in Figures 2 and 3, we obtained rates for TR-driven transitions between $J$-manifolds or even individual spin-rotational sublevels. These rates provided local probe of the radiation environment in which the molecular ion was situated. The rate of TR-driven transition between a given pair of spin-rotational levels can be expressed as $\Gamma_{\mathcal{J},\mathcal{J'}} = \rho_{\textrm{TR}}B_{\mathcal{J},\mathcal{J'}} + A_{\mathcal{J},\mathcal{J'}}$. Here, $\rho_{\textrm{TR}}$ is the energy density of TR, and $A$ and $B$ are the Einstein coefficients for spontaneous emission and absorption/stimulation emission, respectively (SM).
In an ideal blackbody environment, the radiation is randomly polarized and its energy density is given by Planck's law $\rho_{\textrm{BBR}}(\nu, T) = \frac{8\pi h \nu^3}{c^3} \frac{1}{\exp{[h\nu/(k_B T)]} - 1}$.
Here, $\nu$ is the frequency of the radiation and $T$ is the temperature of the environment.
Under the assumption that $\rho_{\textrm{TR}} = \rho_{\textrm{BBR}}$, we derived a value for $T$ from $\Gamma^{\textrm{meas.}}_{J = 0 \rightarrow 1}$ ($\Gamma^{\textrm{meas.}}_{J = 1 \rightarrow 2}$), which is the measured rate of transition between a sublevel in $J = 0$ ($J = 1$) and all allowed sublevels in $J = 1$ ($J = 2$) (Fig. \ref{figBBRCompare}A, upper panel)(SM).
The individual transitions which contributed to these total measured rates are highlighted in the lower panel of Fig. \ref{figBBRCompare}A. With the finite duration required for a molecular state detection, a certain fraction of transition events was not registered, making the measured rates an underestimate of the actual rates.
As such, the derived $T$ should be considered as lower bounds. Nevertheless, we found even the lower estimates of $T$ (400 K and 333 K) to be higher than the ambient temperature of our experiment, which was $\sim$300 K.
Further details about the environment can be obtained by examining transitions between individual sublevels. In particular, we studied the degree to which the 285 GHz frequency component of TR was anisotropic by comparing the rates or probabilities of transitions between sublevels of $J = 0$ and 1 that were driven by different polarizations. Fig. \ref{figBBRCompare}B 
shows the ratios between the energy densities of $\sigma^{-}$- and $\pi$-polarized TR, derived from the measured rates of three pairs of transitions highlighted in the inset (SM). The arrows indicate the direction of the transition.
We observe that the ratios, which were derived from different measurements, were reasonably consistent with each other but larger than unity. Such a result was inconsistent with a randomly polarized field environment where the energy density for every polarization component is equal.
Together, the data from Fig. \ref{figBBRCompare}A and B implied that the thermal environment in the hundreds of GHz spectral region deviated from an ideal blackbody at room temperature.
Possible explanations of this deviation include the elevated temperatures of the electrodes due to the trap RF drive, and the structure of our ion trap.
On the latter point, the trap electrodes might be approximated as a set of conductive planes surrounding the molecular ion (Fig. \ref{figIntro}A). 
Because the radiation components driving rotational transitions (285 and 570 GHz) have wavelengths (1.1 and 0.53 mm) that are longer than the spacing between these conductive planes (0.2 mm), their spectral and polarization characteristics might be substantially modified. 
Modification of the BBR spectrum by a cavity was observed in a previous work, in which sodium Rydberg atoms placed between two parallel plates experience inhibited absorption of TR \cite{vaidyanathan1981inhibited}.
Our results open the possibility of using a molecular ion as an \textit{in-situ} probe of its radiation environment (SM).

The molecular state control protocol demonstrated here is, in principle, generally applicable to heteronuclear molecular ions which are susceptible to TR-driven dynamics (SM). Many proposed precision measurement experiments based on the platform of QLS on single molecular ions \cite{wellers2022controlled, zhou2024quantum, arrowsmith2023opportunities} aim to reach higher spectroscopic accuracy over the current records set by experiments using ensembles of ions \cite{patra2020proton,alighanbari2020precise,roussy2023improved}. For these proposed experiments, improvements in state control would lead to higher data rate and therefore reduced averaging time for spectroscopy.
More broadly, we have demonstrated QLS as a versatile and fully state-resolving tool for single molecule state analysis. When combined with other well-established physical chemistry techniques such as ultrafast lasers and molecular beams which initiate the dynamics, QLS provides an opportunity to detect molecular response to external perturbations including strong-pulse excitation \cite{ohmori2009wave, antonov2021precisely} and inelastic collisions \cite{eberle2015ion, hernandez2017rotationally} on an unprecedented single molecule, single state level.

\bibliography{refs.bib}

\section*{Acknowledgments}
We thank L. Liu and B. Margulis for careful reading of the manuscript. \textbf{Funding:} We acknowledge support from the Army Research Office under Grant W911NF-19-1-0172. \textbf{Author contributions:} Y.L., D.R.L., D.L., and C.-w.C. conceived and designed the experiments. Y.L., Z.L., and C.-w.C. collected and analyzed the data. Y.L. and J.S. set up sources for 285 and 570 GHz microwaves. All authors provided suggestions for the experiments, discussed the results and contributed to editing the manuscript. \textbf{Competing interests:} The authors declare that they have no competing financial interests. \textbf{Data and materials availability:} Data underlying various plots in this article are deposited in the NIST Science Data Portal \cite{nistDataBase2024}. All other data needed to evaluate the conclusions in the paper are present in the paper or the supplementary materials.

\section*{Supplementary materials}
Supplementary Text\\
Figures S1 to S4\\
Tables S1 to S3\\
References \textit{(47 - 50)}

\setcounter{table}{0}
\renewcommand{\thetable}{S\arabic{table}}
\setcounter{figure}{0}
\renewcommand{\thefigure}{S\arabic{figure}}
\renewcommand{\theHfigure}{S\arabic{figure}}
\setcounter{page}{1}
\renewcommand{\thesection}{\large S\arabic{section}}
\renewcommand{\thesubsection}{\normalsize S\arabic{section}.\arabic{subsection}}
\renewcommand{\theequation}{S.\arabic{equation}}


\clearpage
\noindent{\textbf{\large Materials and Methods}}

\bigskip

\noindent{\underline{Molecular state detection, projection, and pumping using quantum-logic spectroscopy}}

In a quantum-logic spectroscopy (QLS) operation, information about the internal state of a spectroscopy ion ($^{40}$CaH$^+$) is coherently transferred onto a logic ion ($^{40}$Ca$^+$) via their shared harmonic normal modes of motion within the ion trap potential. 
First, Ca$^+$ is prepared in the state $| D\rangle \equiv | D_{5/2}, m_a = -5/2 \rangle$ and the axial out-of-phase motional mode in its ground state ($|n = 0 \rangle$). Here, $m_a$ is the magnetic quantum number of Ca$^+$, and $n$ is the harmonic oscillator quantum number for the motion. Then, a sideband transition that attempts to change the internal state of CaH$^+$ while adding a quantum of excitation to the motion ($ |\mathcal J \rangle | 0 \rangle \rightarrow |\mathcal J' \rangle | 1 \rangle $) is driven by a pair of far-detuned 1064 nm laser beams in a Raman configuration. The two beams have $\sigma^-$ and $\pi$ polarizations, respectively, which drives transitions between $|\mathcal J \rangle = |J, m, \xi \rangle $ and $|\mathcal J' \rangle = | J, m\pm1, \xi'\rangle $. Next, a motion-subtracting sideband transition is attempted on Ca$^+$ ($ | D \rangle | 1 \rangle \rightarrow | S \rangle | 0 \rangle $). Here, $| S \rangle \equiv | S_{1/2}, m_a = -1/2 \rangle$ is a bright state under the excitation of a 397 nm laser while repumped by a 866 nm laser, and can be distinguished from the dark state $| D \rangle$ with high fidelity. The detection of atomic fluorescence at the end of the sequence indicates that the molecular transition was successfully driven, which implies detection of the molecule in $|\mathcal J \rangle$. This also non-destructively heralds the molecule's projection into $|\mathcal J' \rangle$. Each projection attempt takes a total of 5.8 ms, which consists of ground-state cooling of the axial out-of-phase mode (4.1 ms), sideband operations on the atomic and molecular ions (1.5 ms), and fluoresence detection on the atomic ion (0.2 ms).

\begin{figure}[t!]
\centering
\includegraphics[width=1.0\textwidth]{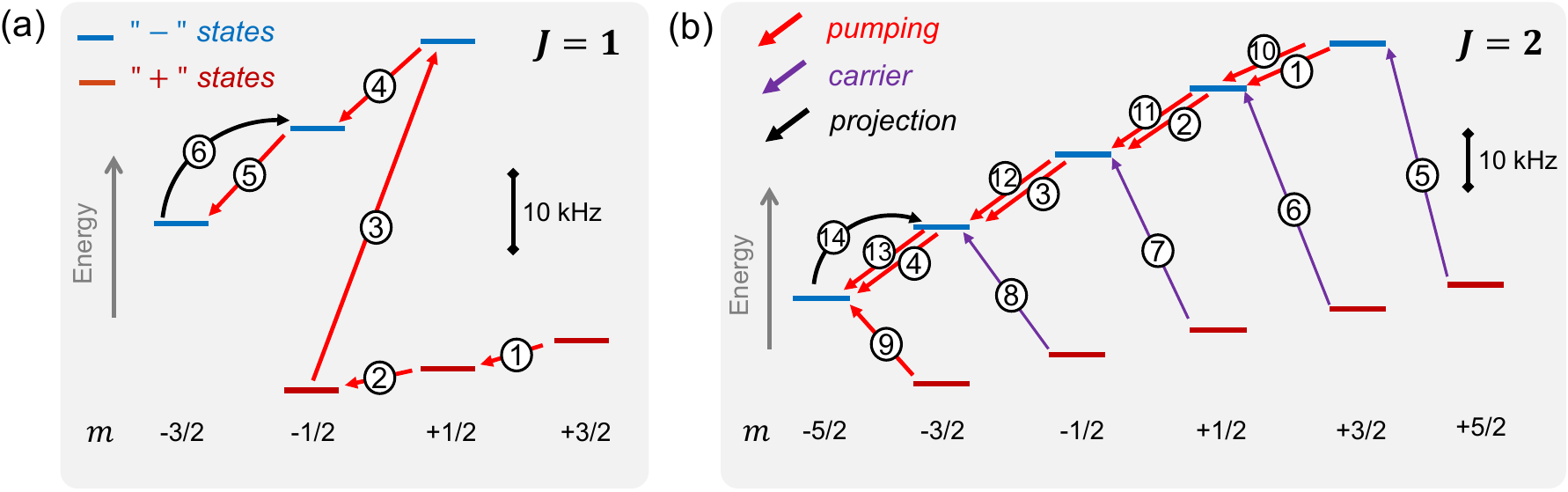}
    \caption{\small``pump \& project" pulse sequences for detecting in $J = 1$ (a) and $J = 2$ (b) manifolds. Red and black arrows represent pumping and projection transitions, respectively. The circled numbers indicate the order in which the transitions are driven. For $J = 2$, we use ``carrier'' transitions (purple arrows), which change only the internal state the molecule but not the motion, to transfer the probabilities from states which suffer from low pumping efficiency to states which can be efficiently pumped. The energy level structures are to scale and calculated for $^{40}$CaH$^+$ in the presence of co-aligned DC magnetic ($B = 6.5 $ G) and RF electric ($E_{\text{RF}} = 1300 $ V/m) fields. A scale bar marking 10 kHz is provided in both (a) and (b).}
\label{J1_2PumpingSequences}
\end{figure}

While QLS in principle allows the detection/projection of the molecule in any state, most Raman transitions in $^{40}$CaH$^+$ are too weak and/or not spectrally resolved from other transitions to allow unambiguous identification of the detected/projected state. The exceptions include transitions between $|J, m = -J - 1/2, - \rangle$ and $|J, m = -J + 1/2, - \rangle$, which are strong and distinctive in frequency from one another and all other Raman transitions \cite{chou2017preparation}, and are referred to as ``signature" transitions. Given these characteristics, we detect the molecular state by first combining the probability of finding the molecule in each sublevel of a given $J$-manifold, $P_{\mathcal{J}}$, into $|J, m = -J - 1/2, - \rangle$ via QLS-based pumping \cite{chou2017preparation}, and then attempt a projection using the signature transition. 
In each pumping step, we drive a $\pi$-pulse on a motion-adding sideband transition $|\mathcal J \rangle = |J, m, \xi \rangle |n = 0\rangle \rightarrow |\mathcal J' \rangle = | J, m\pm1, \xi'\rangle |n = 1\rangle$, and then perform ground-state cooling on the motion ($|n = 1\rangle \rightarrow |n = 0\rangle$).
This introduces dissipation in the system that combines the probabilities in $|\mathcal J \rangle$ into $|\mathcal J' \rangle$. 
This detection method is referred to as ``pump \& project", and the pulse sequences are illustrated schematically in Figure \ref{J1_2PumpingSequences} for $J = 1$ and 2. Their durations are 20.4 and 32.3 ms, respectively. Following the full pumping sequence, the probability in $|J, m = -J - 1/2, - \rangle$ is $\zeta \sum_{\mathcal{J}\in J} P_{\mathcal{J}}$, where $\zeta$ is the overall pumping efficiency. For the illustrated pulse sequences, we measure $\zeta = 0.95(5)$ for $J = 1$ and $0.90(5)$ for $J = 2$. When a pumping sequence similar to that for $J = 2$ is applied to $J \geq 3$, we currently find $\zeta < 0.5$, thus preventing the efficient detection of these manifolds.

\bigskip
\noindent{\underline{State-resolved detection of the $J = 1$ manifold}}

In this section, we describe the procedure used to detect individual states of the $J = 1$ manifold and observe quantum jumps from $J = 0$ to 1 (Fig. \ref{figPopTrackJ0}B).
After preparing the molecule in either sublevel of $J = 0$ and the logic ion and the motion in $|D\rangle |0 \rangle$, we detect each sublevel of $J = 1$ (\textit{e.g.}, $| 1, -3/2, -\rangle$) by projecting it towards a neighboring sublevel (\textit{e.g.}, $| 1, -1/2, -\rangle$). Upon a successful initial projection (\textit{e.g.}, $| 1, -3/2, -\rangle \rightarrow | 1, -1/2, -\rangle$), two additional projections (\textit{e.g.}, $| 1, -1/2, -\rangle \rightarrow | 1, -3/2, -\rangle$ and $| 1, -3/2, -\rangle \rightarrow | 1, -1/2, -\rangle$) are required to confirm the result of the first projection. We observed that the multiple projections suppress the false positive probability to $< 0.2~\%$. Primary sources for false positive events in our system include heating of the axial out-of-phase mode ($|D\rangle |0 \rangle \rightarrow |D\rangle | 1 \rangle$) and spontaneous decay of Ca$^{+}$ ($|D\rangle | 0 \rangle \rightarrow |S\rangle | 0 \rangle$), both of which will result in Ca$^{+}$ in the $|S \rangle$ state after the motion-subtracting atomic sideband and a false positive signal. Without loss of generality, we detect the sublevels in the order $| 1, -3/2, -\rangle$, $| 1, -1/2, +\rangle$, $| 1, -1/2, -\rangle$, $| 1, +1/2, +\rangle$, $| 1, +1/2, -\rangle$, and $| 1, +3/2, +\rangle$.

\begin{figure}[t!]
\centering
\includegraphics[width=0.75\textwidth]{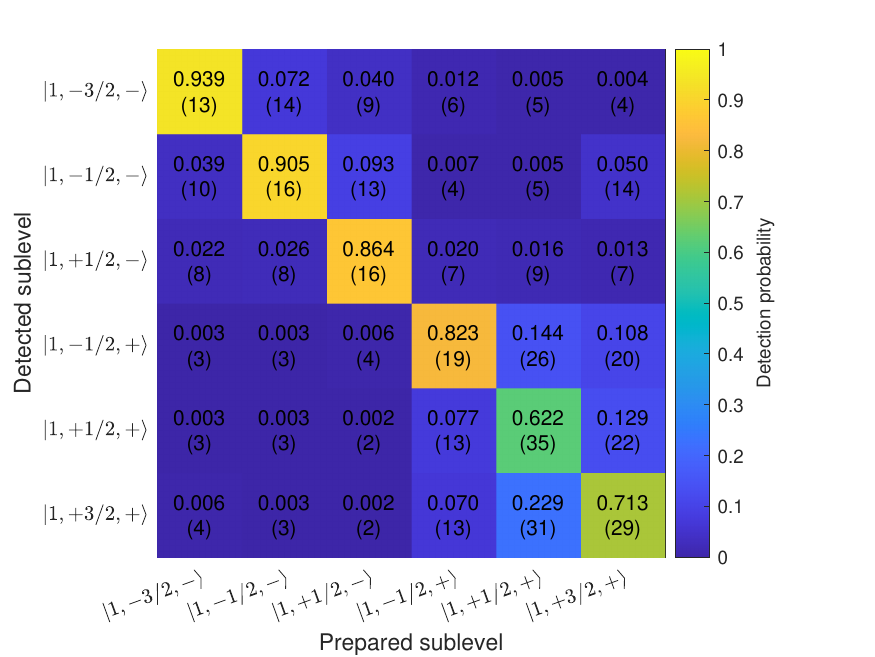}
    \caption{\small Probability of detecting the molecule in each of the 6 sublevels of $J = 1$ following an initial preparation in one of the sublevels. The numbers in the parentheses represent 1SD uncertainty from the measurements.}
\label{J1ReadoutEfficiencies}
\end{figure}

Another source of imperfection during our detection is cross-talk, whereby the occupation of one sublevel is falsely assigned to another one. Contributing factors to cross-talk include the fact that some neighboring sublevels share the same pair of molecular sidebands for detection (\textit{e.g.}, $| 1, -3/2, -\rangle$ and $| 1, -1/2, -\rangle$, $| 1, -1/2, +\rangle$ and $| 1, +1/2, +\rangle$), non-ideal spectral separation between some of the sidebands (especially those within the ``$+$'' manifold), and the requirement for multiple consecutive detections which may shuffle the molecule to a different sublevel than the one it started in. To characterize the effect of cross-talks on the detection, we performed experiments where the molecular ion is deliberately prepared in a given sublevel of $J = 1$, and all 6 sublevels are detected using a sequence identical to that used to monitor quantum jumps from $J = 0$ to 1. The resulting state histograms are displayed in Figure \ref{J1ReadoutEfficiencies}, and are used to convert the raw data into that shown in Fig. \ref{figPopTrackJ0}B. The ``$+$'' states suffer from worse cross-talk due to the smaller frequency separation between the detection sidebands compared to those used for the ``$-$'' states.

\bigskip
\noindent{\textbf{\large Supplementary Text}}

\bigskip
\noindent{\underline{Estimating rates/probabilities for transitions between $J = 0$ and 1}}

In this section, we discuss the calculation of the relative rates (or, equivalently, probabilities) of transitions from either sublevel of $J = 0$ to different sublevels of $J = 1$ by an isotropic, nuclear spin conserving process (\textit{e.g.}, background gas collision, blackbody radiation) (Fig. \ref{figPopTrackJ0}B). Any sublevel of $^{40}$CaH$^+$ $| \mathcal{J} \rangle = |J, m, \xi \rangle$ may be written in the basis $| J, m_J \rangle | I, m_I \rangle$ as
\begin{equation}
    |\mathcal{J} \rangle = c^{-}_{\mathcal{J}} |J, m_J = m + 1/2\rangle | I = 1/2, m_I = -1/2 \rangle + c^{+}_{\mathcal{J}} |J, m_J = m - 1/2\rangle | I = 1/2, m_I = 1/2 \rangle,
\end{equation}
where $c^{-}$ and $c^{+}$ are the state amplitudes for the $ m_I = -1/2$ and $ m_I = 1/2$ components of the state, respectively, and $m_J \in \{ -J, -J + 1, ..., J - 1, J \}$. Consider a transition between two sublevels $\mathcal{J}$ and $\mathcal{J}'$. The transition matrix element can be expressed as $M_{\mathcal{J},\mathcal{J'}} = \langle \mathcal{J} | \mathbf{M} | \mathcal{J'} \rangle$, where $\mathbf{M}$ is the operator corresponding to the process, \textit{e.g.,} dipole operator for TR, potential energy operator for collisions. Assuming $\mathbf{M}$ acts only on the rotational but not the nuclear spin part of the state, we have
\begin{equation} \label{TME}
    M_{\mathcal{J},\mathcal{J'}} = c^{-}_{\mathcal{J}}c^{-}_{\mathcal{J'}} \langle J, m_J = m + 1/2 | \mathbf{M} | J', m_J' = m' + 1/2 \rangle + c^{+}_{\mathcal{J}}c^{+}_{\mathcal{J'}} \langle J, m_J = m - 1/2 | \mathbf{M} | J', m_J' = m' - 1/2 \rangle.
\end{equation}

\begin{table}[t!]
\centering
\addtolength{\leftskip} {-2cm}
\addtolength{\rightskip}{-2cm}
\begin{threeparttable}
\small
\begin{tabular}{lcccccc}
\hline\hline
 & $| 1, -3/2, - \rangle$ & $| 1, -1/2, - \rangle$ & $| 1, 1/2, - \rangle$ & $| 1, -1/2, + \rangle$ & $| 1, 1/2, + \rangle$ & $| 1, 3/2, + \rangle$ \\
\hline
$c^{-}_{|\mathcal{J'}\in J = 1\rangle}$ & 1 & 0.9846 & 0.9894 & 0.1749 & 0.1450 & 0 \\
$P_{|0, -1/2, - \rangle,\mathcal{J'}\in J = 1}$ & 0.3333 & 0.3231 & 0.3263 & 0.0102 & 0.0070 & 0 \\
$c^{+}_{|\mathcal{J'}\in J = 1\rangle}$ & 0& $-0.1749$ & $-0.1450$ & 0.9846& 0.9894& 1 \\
$P_{|0, 1/2, + \rangle,\mathcal{J'}\in J = 1}$ & 0  &  0.0102  &  0.0070  &  0.3231  &  0.3263  &  0.3333 \\
\hline\hline
\end{tabular}
\caption{\small \label{tab:CGSJ0to1} State amplitudes for $J = 1$ sublevels ($c^{\pm}_{\mathcal{J'}\in J = 1}$) and the corresponding probabilities of undergoing a transition from $| 0, \pm1/2 \rangle$ to $J = 1$ sublevels ($P_{|0, \pm1/2, \pm \rangle,|\mathcal{J'}\in J = 1\rangle}$) for an isotropic, nuclear spin conserving process.}
\end{threeparttable}
\end{table}

Now consider transitions between $J = 0$ and 1. Since $c^{-}_{|0, -1/2, - \rangle} = c^{+}_{|0, 1/2, + \rangle} = 1$ and $c^{+}_{|0, -1/2, - \rangle} = c^{-}_{|0, 1/2, + \rangle} = 0$, we have $M_{|0, -1/2, - \rangle, |\mathcal{J'} \in J = 1 \rangle} = c^{-}_{|\mathcal{J'} \in J = 1\rangle} \langle 0, 0 | \mathbf{M} | 1, m_J' \rangle$, and $M_{|0, 1/2, + \rangle,|\mathcal{J'}\in J = 1\rangle} = c^{+}_{|\mathcal{J'}\in J = 1\rangle} \langle 0, 0 | \mathbf{M} | 1, m_J'\rangle$. Without \textit{a priori} knowledge about the details of the process, we assume, for the purpose of this estimate, that it is isotropic. This means that $\langle 0, 0 | \mathbf{M} | 1, m_J'\rangle$ has the same value for $m_j' = -1, 0,$ and 1, and the strength of the process (\textit{e.g.}, number of background gas molecules colliding with CaH$^+$, energy density of the TR field) is equal for the transitions from $m_j = 0$ to $m_j' = -1, 0,$ and 1. In this case, $M_{|0, \pm1/2, \pm \rangle,|\mathcal{J'}\in J = 1\rangle} \propto c^{\pm}_{|\mathcal{J'}\in J = 1\rangle}$, and the corresponding transition rates are $ \Gamma_{|0, \pm1/2, \pm \rangle,|\mathcal{J'}\in J = 1\rangle} \propto |M_{|0, \pm1/2, \pm \rangle,|\mathcal{J'}\in J = 1\rangle}|^2 \propto |c^{\pm}_{|\mathcal{J'}\in J = 1\rangle}|^2$. The rates are in turn proportional to the probabilities of finding the molecule in $| \mathcal{J'} \in J = 1 \rangle$ following its initialization in one of the two $| \mathcal{J} \in J = 0 \rangle $ sublevels, and we calculate these probabilities according to $P_{|0, \pm1/2, \pm \rangle,|\mathcal{J'}\in J = 1\rangle} = | c^{\pm}_{|\mathcal{J'}\in J = 1\rangle} |^2 /\sum_{|\mathcal{J'}\in J = 1\rangle} | c^{\pm}_{|\mathcal{J'}\in J = 1\rangle} |^2 $. Their values are displayed in Tab. \ref{tab:CGSJ0to1} along with the values of $c^{\pm}_{|\mathcal{J'}\in J = 1\rangle}$ from which they derive. The $c^{\pm}_{|\mathcal{J'}\in J = 1\rangle}$ are calculated by diagonalizing the Hamiltonian for CaH$^+$ under an intentionally applied external DC magnetic field of 6.5 G, and a co-aligned residual RF electric field of 1300 V/m in amplitude which cannot be compensated \cite{collopy2023effects}.

\bigskip

\noindent{\underline{Rates for transitions driven by thermal radiation}}

The TR-driven transition rate between two individual sublevels $| \mathcal{J} \rangle = | J, m, \xi \rangle$ and $| \mathcal{J'} \rangle = | J', m', \xi' \rangle$ is given by 
\begin{equation} \label{Gamma_TR}
    \Gamma_{\mathcal{J} \rightarrow \mathcal{J'}} = \rho^\textrm{TR}_{\mathcal{J},\mathcal{J'}} B_{\mathcal{J},\mathcal{J'}} + A_{\mathcal{J},\mathcal{J'}},    
\end{equation}
where $\rho_\textrm{TR}$ is energy density for the component of TR driving the $| \mathcal{J} \rangle \rightarrow |\mathcal{J'} \rangle$ transition, $B_{\mathcal{J},\mathcal{J'}} = \frac{2 \pi^2}{3 h^2 \epsilon_0} |\mu_{\mathcal{J},\mathcal{J'}}|^2 $ is the Einstein coefficient for stimulated absorption or emission, and $A_{\mathcal{J},\mathcal{J'}} = \frac{16 \pi^3 \nu^3}{3 h \epsilon_0 c^3} |\mu_{\mathcal{J},\mathcal{J'}}|^2 $ is the Einstein coefficient for spontaneous emission, which is only nonzero for $J' < J$. $\mu_{\mathcal{J},\mathcal{J'}}$ is given by Eq. \ref{TME}, in which the generic operator $\mathbf{M}$ is replaced by the dipole operator $\vec{\mu}$. We can evaluate $\mu_{\mathcal{J},\mathcal{J'}}$ using state amplitudes obtained from solving the molecular Hamiltonian, the previously measured ground state permanent dipole moment of $^{40}$CaH$^+$ $\mu_{\textrm{CaH}^+}$ = 5.34(19) Debye \cite{collopy2023effects}, as well as $\langle J, m_J | \vec{\mu} | J', m_J' \rangle$ calculated using formulas for the matrix elements of pure rotational transitions (Ref. \cite{townes2013microwave} p. 22). According to dipole selection rules, $\langle J, m_J | \vec{\mu} | J', m_J' \rangle \neq 0$ for $J' = J \pm 1$ and $m_J' = m_J, m_J\pm 1$.

Using Eq. \ref{Gamma_TR}, we can estimate the rates for TR-driven transitions given a specific model for $\rho_\textrm{TR}$, \textit{e.g.,} Planck's law for blackbody radiation $\rho_{\textrm{BBR}}(\nu, T) = \frac{8\pi h \nu^3}{c^3} \frac{1}{\exp{[h\nu/(k_B T)]} - 1}$. To obtain comparisons for the measured rates $\Gamma^{\textrm{meas.}}_{J = 0 \rightarrow 1}$ and $\Gamma^{\textrm{meas.}}_{J = 1 \rightarrow 2}$ (Fig. \ref{figBBRCompare}A, upper panel), we calculate $\Gamma^{\textrm{BBR}}_{J = 0 \rightarrow 1}$ and $\Gamma^{\textrm{BBR}}_{J = 1 \rightarrow 2}$ by setting $\rho_{\textrm{TR}} = \rho_{\textrm{BBR}}$ and summing over the transition rates from the relevant sublevels (Fig. \ref{figBBRCompare}A, lower panel).
For either $| \mathcal{J} \in J = 0 \rangle $, we find $\Gamma^{\textrm{BBR}}_{J = 0 \rightarrow 1} = \rho_\textrm{BBR}(\nu = 285~\textrm{GHz}, T) \frac{2 \pi^2}{3 h^2 \epsilon_0} \sum_{\mathcal{J'}\in J = 1} |\mu_{\mathcal{J}, \mathcal{J'} }|^2 $, where $ \sum_{\mathcal{J'}\in J = 1} |\mu_{\mathcal{J}, \mathcal{J'} }|^2 = \mu^2_{\textrm{CaH}^+} $. Similarly, for any $| \mathcal{J} \in J = 1 \rangle $, we find $\Gamma^{\textrm{BBR}}_{J = 1 \rightarrow 2} = \rho_\textrm{BBR}(\nu = 570~\textrm{GHz}, T) \frac{2 \pi^2}{3 h^2 \epsilon_0} \sum_{\mathcal{J'}\in J = 2} |\mu_{\mathcal{J}, \mathcal{J'} }|^2 $, where $ \sum_{\mathcal{J'}\in J = 2} |\mu_{\mathcal{J}, \mathcal{J'} }|^2 = \frac{2}{3} \mu^2_{\textrm{CaH}^+} $. The uncertainty in the measured $\mu_{\textrm{CaH}^+}$ propagates into the calculated values for $\Gamma^{\textrm{BBR}}_{J = 0 \rightarrow 1}$ and $\Gamma^{\textrm{BBR}}_{J = 1 \rightarrow 2}$. 
By comparing $\Gamma^{\textrm{meas.}}$ and $\Gamma^{\textrm{BBR}}$ along with their associated uncertainties, we find the range of values for $T$ over which the blackbody model agrees with our measurements (Fig. \ref{figBBRCompare}A, upper panel).

We may also use Eq. \ref{Gamma_TR} to derive the energy density of TR from measured transition rates or probabilities.
We do so to obtain the ratios between the energy densities of $\sigma^-$- and $\pi$-polarized components of TR at $\sim$ 285 GHz, ${\rho^\textrm{TR}_{\sigma^-}}/{\rho^\textrm{TR}_{\pi}}$, which are presented in Fig. \ref{figBBRCompare}B. For $J = 0 \rightarrow 1$ transitions, the term $A_{\mathcal{J},\mathcal{J'}}$ is zero, and we have
\begin{equation*}
    \frac{\rho^\textrm{TR}_{\mathcal{J} \in J = 0,\mathcal{J'} \in J = 1}}{\rho^\textrm{TR}_{\mathcal{J} \in J = 0,\mathcal{J''} \in J = 1}} = \frac{\Gamma^{\textrm{meas.}}_{\mathcal{J} \in J = 0,\mathcal{J'} \in J = 1}}{\Gamma^{\textrm{meas.}}_{\mathcal{J} \in J = 0,\mathcal{J''} \in J = 1}} \frac{|\mu_{\mathcal{J} \in J = 0,\mathcal{J''} \in J = 1}|^2}{|\mu_{\mathcal{J} \in J = 0,\mathcal{J'} \in J = 1}|^2}.
\end{equation*}
 Using the transition probabilities presented in Fig. \ref{figPopTrackJ0}B (which are proportional to rates) and the state amplitudes from Tab. \ref{tab:CGSJ0to1}, we calculate the energy density ratios for the cases $\{|\mathcal{J}\rangle = |0, -1/2,-\rangle\}, |\mathcal{J'}\rangle = |1, -3/2, -\rangle, |\mathcal{J''\rangle} = |1, -1/2, -\rangle\}$ and $\{|\mathcal{J}\rangle = |0, 1/2,+\rangle, |\mathcal{J'}\rangle = |1, -1/2, +\rangle, |\mathcal{J''}\rangle = |1, 1/2, +\rangle\}$, which correspond to the purple and green arrow pairs in the inset of Fig. \ref{figBBRCompare}B, respectively. For $J = 1 \rightarrow 0$ transitions, we use  Eq. \ref{Gamma_TR} to convert the measured rates for the transitions $|1,-3/2,-\rangle\rightarrow|0,-1/2,-\rangle$ and $|1,-1/2,-\rangle\rightarrow|0,-1/2,-\rangle$ (blue arrows in Fig. \ref{figBBRCompare}B inset) into corresponding energy densities for $\sigma^-$- and $\pi$-polarized TR, and then take their ratio.
The rates for these transitions are obtained from the molecular state control experiments summarized in Fig. \ref{figPopTrackJ1}. The molecule is prepared in either $|1,-3/2,-\rangle$ or $|1,-1/2,-\rangle$, the number of transitions to $|0,-1/2,-\rangle$ is recorded over the duration of the sequence, and the rate is calculated by dividing the number of transitions by the time the molecule spent in the prepared $J = 1$ sublevel.
The rate data are collected over three sets on different days, yielding the three blue square data points in Fig. \ref{figBBRCompare}B. All data used to calculate the energy density ratios are presented in Tab. \ref{tab:energyDensityRatio}.

\begin{table} [t!]
\centering
\addtolength{\leftskip} {-2cm}
\addtolength{\rightskip}{-2cm}
\begin{threeparttable}
\small
\begin{tabular}{cccccc}
\hline\hline
 Data type & $\sigma^-$ transition & value & $\pi$ transition & value & $\rho^\textrm{TR}_{\sigma^-}/{\rho^\textrm{TR}_{\pi}}$ \\
\hline
Probability & $|0, -1/2, -\rangle \rightarrow |1, -3/2, -\rangle$ & 0.342(17) & $|0, -1/2, -\rangle \rightarrow |1, -1/2, -\rangle$ & 0.259(17) & 1.28(10) \\
Probability & $|0, 1/2, +\rangle \rightarrow |1, -1/2, +\rangle$ & 0.358(17) & $|0, 1/2, +\rangle \rightarrow |1, 1/2, +\rangle$ & 0.299(23) & 1.21(11) \\
Rate        & $|1, -3/2, -\rangle \rightarrow |0, -1/2, -\rangle$ & 0.0799(77) s$^{-1}$ & $|0, -1/2, -\rangle \rightarrow |0, -1/2, -\rangle$ & 0.0472(66) s$^{-1}$ & 1.68(30) \\
      &  & 0.0828(81) s$^{-1}$ &   & 0.0579(61) s$^{-1}$ & 1.41(21) \\
      &   & 0.0802(36) s$^{-1}$ &   & 0.0507(36) s$^{-1}$ & 1.57(14) \\
\hline\hline
\end{tabular}
\caption{\small \label{tab:energyDensityRatio} Transition probability and rate data used to calculate the ratio between $\sigma^-$- and $\pi$-polarized TR presented in Fig. \ref{figBBRCompare}B. All error bars represent 1 SD uncertainty.}
\end{threeparttable}
\end{table}

Molecules possess a variety of different transitions and can be used to sense different parts of the thermal radiation frequency spectrum, and even different polarization components. Similar to atom-based sensors, molecules have the potential for high-precision sensing with the core element of each sensor identically constructed by nature. Such a sensor can be in-situ probes for the thermal environment for precision measurements based on neutral and charged atoms, such as optical atomic clocks.

To accurately measure the properties of thermal radiation with this scheme, the sampling rate for the molecular state needs to be much higher than the transition rates so the jumps between molecular states are all detected. Other mechanisms that can cause such jumps will also need to be characterized to constrain their effects on the measurement. These pose limitations on the dynamic range.

\bigskip
\noindent{\underline{TR-driven transitions to vibrationally excited states}}

While we are primarily concerned with TR-driven transitions between rotational manifolds within the ground vibrational state ($v = 0$) in our state tracking experiments, there exists finite rates of transitions to vibrationally excited states ($v'$). In this section, we estimate these rates from first principles. Specifically, we are interested in the rates of TR-driven transitions from the target manifold of our tracking protocol $\{v = 0, J = 1\}$ to all allowed excited rovibrational manifolds, which are limited by the dipole selection rule to $\{v', J' = 0\}$ and $\{v', J' = 2\}$. For the purpose of this estimate, we work under the assumption that the thermal environment is described by an ideal blackbody at 300 K.

To facilitate the discussions of rovibrational transitions, we label each state as $| v, J, m, \xi \rangle$, where $v$ is the vibrational quantum number. Due to dipole selection rules of TR-driven transitions, $| v = 0, J = 1, m, \xi \rangle$ can only couple to $| v', J', m', \xi'\rangle$ for which $J' - J = \pm 1$ and $m' - m = 0, \pm 1$.
The transition dipole moment (TDM) of a rovibrational transition is given by the expression \cite{simons1997quantum}
\begin{equation}
    \mu_{| v, J, m, \xi \rangle \rightarrow | v', J', m', \xi' \rangle } = \langle v, J, m, \xi | \vec{\mu} | v', J', m', \xi' \rangle = \langle v | \mu | v' \rangle \langle J, m_J, \xi | \hat{\mu} | J', m_J', \xi' \rangle \\
\end{equation}
where $\vec{\mu}$ is the (electronically averaged) dipole moment operator, $\mu$ is its magnitude along the direction of the Ca-H bond, and $\hat{\mu}$ is a unit vector specifying the orientation of the bond (\textit{i.e.}, $\vec{\mu} = \mu\cdot \hat{\mu}$). $\mu_{v,v'} = \langle v | \mu | v' \rangle$ is the vibrational dipole matrix element, whose diagonal elements represent the permanent dipole moment (PDM) of CaH$^+$ in various vibrational states, and off diagonal elements represent the TDMs between vibrational states. $\hat{\mu}_{\{J, m_J, \xi\},\{J', m_J', \xi'\}} = \langle J, m_J, \xi | \hat{\mu} | J', m_J', \xi' \rangle$ is the rotational factor whose value is, to first order, independent of the vibrational state.

\begin{table}[t!]
\centering
\addtolength{\leftskip} {-2.5cm}
\addtolength{\rightskip}{-2cm}
\begin{threeparttable}
\small
\begin{tabular}{cccccccc}
\hline\hline
 $v'$ & $\mu_{v,v'}$ & $\nu_{\{0,1\}\rightarrow\{v',0\}}$ & $\rho(\nu_{\{0,1\}\rightarrow\{v',0\}}, 300~\text{K})$ & $\Gamma_{\{0,1\}\rightarrow\{v',0\}}$ & $\nu_{\{0,1\}\rightarrow\{v',2\}}$ & $\rho(\nu_{\{0,1\}\rightarrow\{v',2\}}, 300~\text{K})$ & $\Gamma_{\{0,1\}\rightarrow\{v',2\}}$ \\
 ~ & ($e \cdot a_0$) & (THz) & (J m$^{-3}$ Hz$^{-1}$) & (s$^{-1}$) & (THz) & (J m$^{-3}$ Hz$^{-1}$) & (s$^{-1}$) \\ 
\hline
 0 & 2.09 & 0.285 & $3.07\times10^{-22}$ & 0.0549 & 0.570 & $1.20\times10^{-21}$ & 0.429 \\
 1 & $6.70\times10^{-2}$  & 44.2 & $4.66\times10^{-20}$ & $8.48\times10^{-3}$ & 44.8 & $4.31\times10^{-20}$ & 0.0157 \\
 2 & $2.01\times10^{-2}$ & 87.1 & $3.76\times10^{-22}$ & $6.16\times10^{-6}$ & 87.7 & $3.38\times10^{-22}$ & $1.11\times10^{-5}$ \\
 3 & $4.02\times10^{-3}$ & 129 & $1.57\times10^{-24}$ & $1.02\times10^{-9}$ & 129 & $1.39\times10^{-24}$ & $1.82\times10^{-9}$ \\
 4 & $1.13\times10^{-3}$ & 169 & $5.64\times10^{-27}$ & $2.91\times10^{-13}$ & 170 & $5.00\times10^{-27}$ & $5.16\times10^{-13}$ \\
 5 & $3.62\times10^{-4}$ & 208 & $2.09\times10^{-29}$ & $1.11\times10^{-16}$ & 208 & $1.84\times10^{-29}$ & $1.95\times10^{-16}$ \\
\hline\hline
\end{tabular}
\caption{\small \label{tab:rovibTDM} Rates of rovibrational transitions and the parameters used to calculate them. $\mu_{v,v'}$ vibrational dipole matrix element (values are from \textit{ab initio} calculations performed by P. N. Plessow (private communication), which are within $\sim10\%$ of those from Ref. \cite{abe2010ab}); $\nu$: rovibrational transition frequency; $\rho$: TR energy density (BBR model at 300 K); $\Gamma$: transition rate.}
\end{threeparttable}
\end{table}

We use Eq. \ref{Gamma_TR} to calculate the rates of rovibrational transitions driven by TR, from a sublevel of $\{ v = 0, J = 1 \}$ to all sublevels of $\{ v', J' = 0, 2\}$ (the effects of rovibrational transitions from $\{ v = 0, J = 0 \}$ and $\{ v = 0, J = 2 \}$ can be ignored, since the molecule are only found in these manifolds during recoveries, and the recoveries account for $\lesssim 4\%$ of the total duration where the state is tracked). We suppress the quantum numbers $m$ and $\xi$ in state labeling, since the rate is identical for any sublevel of $\{ v = 0, J = 1 \}$.
The calculated rates, along with the dipole moments and TR energy densities used in the calculations, are presented in Tab. \ref{tab:rovibTDM}. At 300 K, non-negligible vibrational transitions occurs only to $v' = 1$. These transitions result in leakage in the probability to find the molecule in the tracked subspace and impacts the performance of our state tracking protocol, which we discuss in a following section.

\bigskip
\noindent{\underline{State tracking experiment timing}}

\begin{figure}[t!]
\centering
\includegraphics[width=1.00\textwidth]{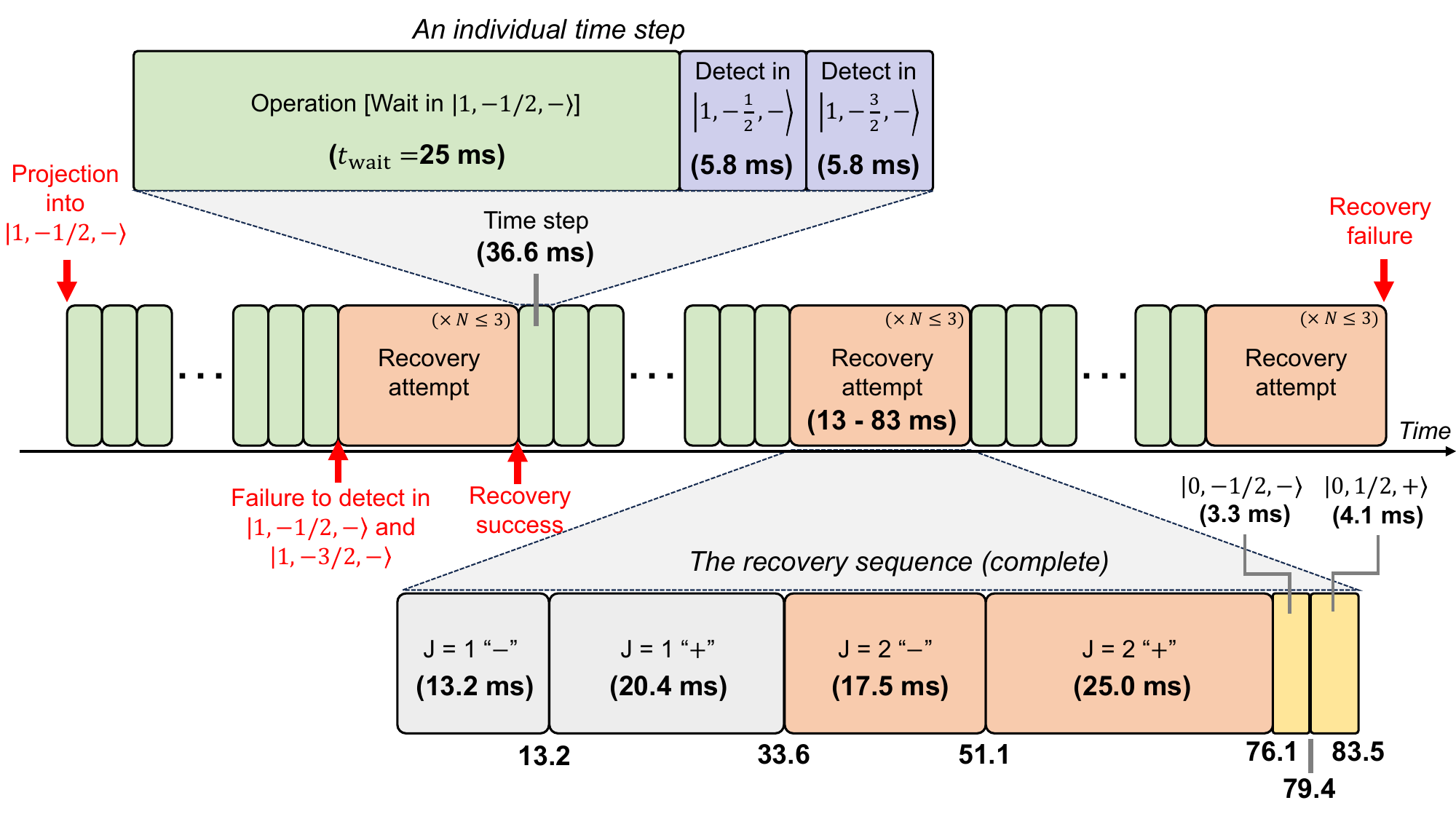}
    \caption{\small \textbf{Timing of the state tracking experiment}. Center panel: an example of experiment timing during a period of continuous state tracking beginning with projection of the state into $| 1, -1/2, - \rangle$ and ends with the failure to recovery the state back to $J = 1$. Upper panel: an individual time step which consists of a waiting period for coherent operations and two QLS detections from $| 1, -1/2, - \rangle$ and $| 1, -3/2, - \rangle$, respectively. Lower panel: the recovery sequence, in which the state is searched in four sub-manifolds: the ``$-$" states of $J = 1$, the ``$+$" states of $J = 1$, the ``$-$" states of $J = 2$, the ``$+$" states of $J = 2$, and finally the two sublevels of $J = 0$. }
\label{stateTrackingTiming}
\end{figure}

In this section we present the timing of various procedures during the state tracking and control experiment presented in Fig. \ref{figPopTrackJ1}. A schematic timing diagram of the experimental sequence is presented in Figure \ref{stateTrackingTiming}. The center panel represents a period of continuous state tracking, which begins with projection of the state into $| \mathcal{J}_i \rangle = | 1, -1/2, - \rangle$ and ends with the failure to recovery the state back to $J = 1$. This corresponds to a green block from Fig. \ref{figPopTrackJ1}D. The majority of the sequence consists of a repetition of individual time steps, the details of which are shown in the upper panel of Figure \ref{stateTrackingTiming}. Each individual time step consists of a waiting period of 25 ms (reserved for coherent operations on the molecular state in future experiments), and two QLS detections from $| 1, -1/2, - \rangle$ and $| 1, -3/2, - \rangle$, respectively. The repetition of the time steps is interrupted by a failure to detect in both of these sublevels, which triggers the execution of a sequence of pulses to attempt recovery of the state. The recovery sequence is shown in the lower panel of Figure \ref{stateTrackingTiming}. The sequence uses the ``pump \& project" method to search for the state in four sub-manifolds: the ``$-$" states of $J = 1$ (using pulses \textcircled{4}, \textcircled{5}, and \textcircled{6} in Figure \ref{J1_2PumpingSequences}(a)), the ``$+$" states of $J = 1$ (using pulses \textcircled{1} -- \textcircled{6} in Figure \ref{J1_2PumpingSequences}(a)), the ``$-$" states of $J = 2$ (using pulses \textcircled{1} -- \textcircled{4} and \textcircled{14} in Figure \ref{J1_2PumpingSequences}(b)), and the ``$+$" states of $J = 2$ (using pulses \textcircled{5} -- \textcircled{14} in Figure \ref{J1_2PumpingSequences}(b)). Positive detections in $J = 1$ will automatically resets the state to $| 1, -1/2, - \rangle$ by design of the pulse sequence, while positive detection in $J = 2$ will be followed by a $\sim$570 GHz microwave $\pi$-pulse to drive a $J = 2\rightarrow1$ transition. Additionally, the sequence detects the two sublevels of $J = 0$ by coherently transferring any probability in $|0, -1/2, - \rangle$ ($|0, 1/2, + \rangle$) to $| 1, -3/2, - \rangle$ ($| 1, -1/2, - \rangle$) via a $\sim$285 GHz microwave $\pi$-pulse and then attempting a projection. If the state is recovered at the end of any of the above stages, the recovery sequence terminates and the experiment resumes repetition of the individual time steps. We allow the execution of up to three complete recovery sequences before declaring failure to recover the state, at which point the system goes back to waiting for a projection in the subspace $J = \{0,1,2\}$.

\bigskip
\noindent{\underline{Limitations on state tracking performance}}

\begin{figure}[t!]
\centering
\includegraphics[width=1.00\textwidth]{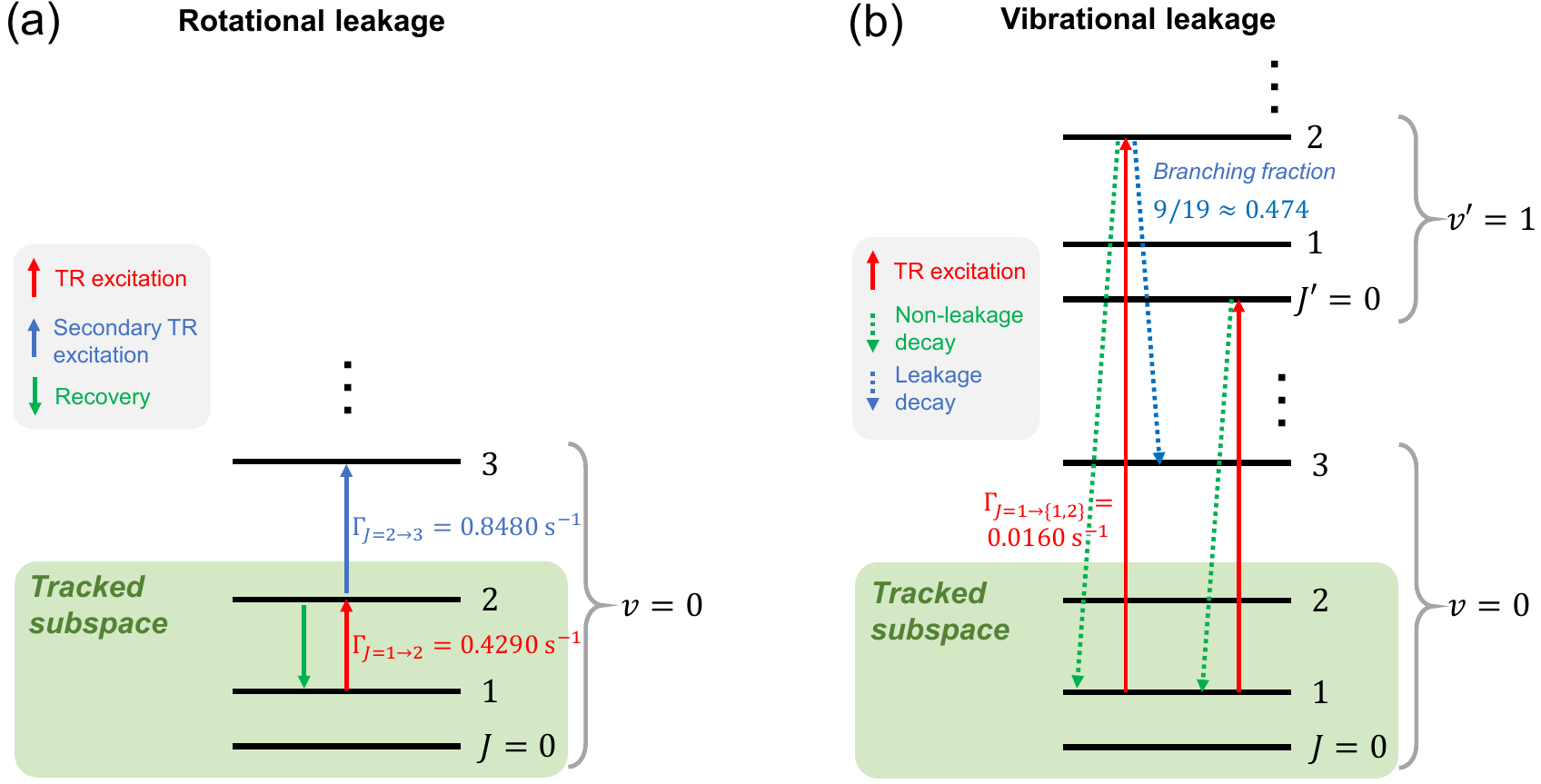}
    \caption{\small \textbf{Leakages from tracked subspace during state tracking}. (a) Rotational leakage: the molecule undergoes TR-driven transitions (red and blue solid arrows) via the path $J = 1 \xrightarrow{\text{TR}} J = 2 \xrightarrow{\text{TR}} J = 3$. This can occur either during an individual time step of the experiment (a second order process), or during the attempt to recover the state from $J = 2$ back to $J = 1$ (a first order process).
    (b) Vibrational leakage: undergoes TR-driven transitions (red solid arrows) from $J = 1$ to the vibrationally excited manifolds $\{v' = 1, J' = 0\}$ and $\{v' = 1, J' = 2\}$. From both manifolds, the molecule can decay (via spontaneous emission) back to $J = 1$ (green dotted arrows), which does not result in a leakage. From $\{v' = 1, J' = 2\}$, the molecule can also decay to $J = 3$ (blue dotted arrows), resulting in a leakage. The branching fraction for the leakage decay ($ \{ v = 1, J = 2 \} \xrightarrow[\text{decay}]{\text{Spont.}} J = 3$) is $9/19$. The rates ($\Gamma$) of the relevant TR-driven transitions are labeled next to the arrows indicating the transitions.
    }
\label{stateTrackingLeakage}
\end{figure}

The performance of our state tracking and control protocol (Fig. \ref{figPopTrackJ1}D) can be measured in terms of the the average duration by which we can continuously keep the molecule in $J = 1$  ($\overline{T}_{\text{track}}$). This duration is limited by processes which cause transitions (``leakage'') out of the tracked subspace $J \in \{0, 1, 2\}$. In this section, we discuss the leakage channels and estimate their rates. We work under the assumption of a 300 K blackbody thermal environment, which approximately describes our system. Given the molecular structure of CaH$^+$, we identify two leakage channels: rotational leakage and vibrational leakage (Figure \ref{stateTrackingLeakage}). We label each  rotational manifold in the ground vibrational state as $J$, and those in the excited vibrational states as $\{ v, J \}$, where $v$ is the vibrational quantum number. For both channels, we consider the leakage to be irreversible once the molecule reaches the $ J = 3 $ manifold. For rotational leakage, the dominant path is $J = 1 \xrightarrow{\text{TR}} J = 2 \xrightarrow{\text{TR}} J = 3$. For vibrational leakage, the dominant path is $J = 1 \xrightarrow{\text{TR}} \{ v = 1, J = 2 \} \xrightarrow[\text{decay}]{\text{Spont.}} J = 3$.

To estimate the leakage rate due to these channels, we first consider the probabilities in various relevant manifolds coupled to $J = 1$ by TR after an individual time step ($\Delta t$) during the tracking experiment. Each time step consists of the wait time, $t_{\text{wait}}$, and the time to check whether the state is still in $J = 1$, $t_{J = 1, \text{check}}$. $t_{J = 1, \text{check}}$ consists primarily of the time to detect in the sublevels $| 1, -1/2, - \rangle$ and $| 1, -3/2, - \rangle$ (5.8 ms each), but also the time for the occasional recoveries from the $J = 1$ ``$-$" states and $J = 1$ ``$+$" states as part of the recovery sequence. The latter process adds, on average, $\sim 1$ ms to each time step (Figure \ref{stateTrackingTiming}, upper panel). The relevant manifolds are $J = 3$, as well as those that can additionally couple to $J = 3$, which include $J = 2$ and $\{ v = 1, J = 2 \}$. In the beginning of a time step, the molecule is initialized in $J = 1$. We can solve for the probabilities using the set of approximate rate equations
\begin{equation} \label{rateEqns1}
    \begin{split}
        \dot{P}_{J = 2}(t) &= \Gamma_{J = 1\rightarrow2} P_{J = 1}(t) \\
        \dot{P}_{J = 3}(t) &= \Gamma_{J = 2\rightarrow3} P_{J = 2}(t) \\
        \dot{P}_{\{v = 1, J = 2\}}(t) &= \Gamma_{\{0,1\} \rightarrow \{1,2\}} P_{J = 1}(t), \\
    \end{split}
\end{equation}
with the initial condition 
$\{P_{J = 1}(0) = 1,~P_{J = 2}(0) = 0,~P_{J = 3}(0) = 0,~P_{\{v = 1, J = 2\}}(0) = 0\}$.
Note that here we have ignored any back flow of probability to $J = 1$ based on the criteria that each time step is very short compared to the inverses of the relevant rates.
To the lowest order, the solutions to Eq. \ref{rateEqns1} at the end of each time step ($t = \Delta t$) are

\begin{equation} \label{rateEqns2}
    \begin{split}
        \Delta P_{J = 2} = P_{J = 2}(\Delta t) &= \Gamma_{J = 1\rightarrow2} \Delta t \\
        \Delta P_{J = 3} = P_{J = 3}(\Delta t) &= \frac{1}{2}\Gamma_{J = 1\rightarrow2}\Gamma_{J = 2\rightarrow3} \Delta t^2 \\
        \Delta P_{\{v = 1, J = 2\}} = P_{\{v = 1, J = 2\}}(\Delta t) &= \Gamma_{\{0,1\} \rightarrow \{1,2\}} \Delta t. \\
    \end{split}
\end{equation}

Now we consider the probability of additional loss from $J = 2$ and $\{ v = 1, J = 2 \}$ to $J = 3$. For $J = 2$, loss to $J = 3$ can occur due to TR-driven transition during the recovery, and the probability is, to first order, $\Gamma_{J = 1\rightarrow3} t_{J = 2,\text{rec.}}$. Here, $t_{J = 2,\text{rec.}}$ represents the average time the recovery sequence takes to bring the molecule back to $J = 1$. Note that $t_{J = 2,\text{rec.}}$ is not deterministic as more than one recovery attempt is occasionally required to reset the state back to $J = 1$. We therefore use the average value of $J = 2$ recovery times measured during the tracking experiments for our estimates. For $\{ v = 1, J = 2 \}$, there is a branching fraction of $R_{\{1, 2 \}\rightarrow J = 3}$ of spontaneous decays to $J = 3$, while the rest of the decays are back to $J = 1$. As such, the total loss probability from $J = 1$ to $J = 3$ after a time step is

\begin{equation}  \label{lossPerTimeStep}
    \begin{split}
        \Delta P_{\text{loss}} &= (\Gamma_{J = 2\rightarrow3} t_{J = 2,\text{rec.}}) \Delta P_{J = 2} + \Delta P_{J = 3} + R_{\{1, 2 \}\rightarrow J = 3} \Delta P_{\{v = 1, J = 2\}} \\
        &= \Gamma_{J = 1\rightarrow2} t_{J = 2,\text{rec.}} \Gamma_{J = 2\rightarrow3}\Delta t + \frac{1}{2}\Gamma_{J = 1\rightarrow2}\Gamma_{J = 2\rightarrow3} \Delta t^2 + R_{\{1, 2 \}\rightarrow J = 3} \Gamma_{\{0,1\} \rightarrow \{1,2\}} \Delta t
    \end{split}
\end{equation}

After each successful recovery the initial conditions of Eq. \ref{rateEqns1} are restored and the protocol proceeds with the next time step. The probability for the loss to occur in the $n$-th time steps is
\[
P_{\text{loss}}(n) = (1 - \Delta P_{\text{loss}})^{n - 1}\Delta P_{\text{loss}},
\]
and the average step at which the loss occurs is
\[
\langle n \rangle = \sum_{n = 1}^{\infty} n P_{\text{loss}}(n) = \frac{1}{\Delta P_{\text{loss}}}.
\]
Therefore the average time for continuous tracking is
\begin{equation}
    \begin{split}
        \overline{T}_{\text{track}} = \langle n \rangle \Delta t &= \frac{\Delta t}{\Gamma_{J = 1\rightarrow2} \Delta t~\Gamma_{J = 2\rightarrow3} t_{J = 2,\text{rec.}} + \frac{1}{2}\Gamma_{J = 1\rightarrow2}\Gamma_{J = 2\rightarrow3} \Delta t^2 + R_{\text{vib.}\rightarrow J = 3} \Gamma_{\text{vib.}} \Delta t} \\
        &= \left(\Gamma_{J = 1\rightarrow2} \Gamma_{J = 2\rightarrow3} t_{J = 2,\text{rec.}} + \frac{1}{2}\Gamma_{J = 1\rightarrow2}\Gamma_{J = 2\rightarrow3} \Delta t + R_{\text{vib.}\rightarrow J = 3} \Gamma_{\text{vib.}}\right)^{-1} \\
        &= \left[\Gamma_{J = 1\rightarrow2} \Gamma_{J = 2\rightarrow3} \left(t_{J = 2,\text{rec.}} + \frac{\Delta t}{2}\right) + R_{\{1, 2 \}\rightarrow J = 3} \Gamma_{\{0,1\} \rightarrow \{1,2\}}\right]^{-1},
    \end{split}
\end{equation}
and the effective rate of loss from $J = 1$ during tracking is
\begin{equation} \label{GammaLoss}
        \Gamma_{\text{loss}} = \overline{T}^{-1}_{\text{track}} = \Gamma_{J = 1\rightarrow2} \Gamma_{J = 1\rightarrow3} \left(t_{J = 2,\text{rec.}} + \frac{\Delta t}{2}\right) + R_{\{1, 2 \}\rightarrow J = 3} \Gamma_{\{0,1\} \rightarrow \{1,2\}} = \Gamma_{\text{loss}}^{\text{rot.}} + \Gamma_{\text{loss}}^{\text{vib.}}.
\end{equation}
Here, $\Gamma_{\text{loss}}^{\text{rot.}} = \Gamma_{J = 1\rightarrow2} \Gamma_{J = 1\rightarrow3} \left(t_{J = 2,\text{rec.}} + \Delta t/2\right)$ represents rotational leakage, and \\$\Gamma_{\text{loss}}^{\text{vib.}} = R_{\{1, 2 \}\rightarrow J = 3} \Gamma_{\{0,1\} \rightarrow \{1,2\}}$ represents vibrational leakage. For the tracking experiment shown in Fig. \ref{figPopTrackJ1}, we estimate $\Gamma_{\text{loss}}^{\text{rot.}} = 0.0268$ s$^{-1}$,  $\Gamma_{\text{loss}}^{\text{vib.}} = 0.0074$ s$^{-1}$, and $\Gamma_{\text{loss}} = 0.0343$  s$^{-1}$, using the calculated rates ($\Gamma_{J = 1 \rightarrow 2} = 0.4290$ s$^{-1}$, $\Gamma_{J = 2 \rightarrow 3} = 0.8480$ s$^{-1}$, $\Gamma_{\{0,1\} \rightarrow \{1,2\}} = 0.0157$ s$^{-1}$), the calculated branching fraction ($R_{\{1, 2 \}\rightarrow J = 3} = 9/19 \approx 0.474$), and the known times from the tracking protocol ($t_{\text{wait}} = 25$ ms, $t_{J = 1, \text{check}} = 12.56$ ms, $\Delta t = t_{\text{wait}} + t_{J = 1, \text{check}} = 37.56$ ms, $t_{J = 2, \text{rec.}} = 55.04$ ms). The total loss rate corresponds to an average continuous tracking time of $\overline{T}_{\text{track}} = \Gamma^{-1}_{\text{loss}} = $ 29.1 s, comparable to the experimental value of 35.5 s. Eq. \ref{GammaLoss} shows that, to the lowest orders, $\Gamma_{\text{loss}}$ has a linear dependence on the wait time for coherent operations (Recall that $\Delta t = t_{\text{wait}} + t_{J = 1, \text{check}}$).

Finally, we note that both rotational and vibrational leakages populate the state $J = 3$. As such, further improvements in the state tracking duration can be achieved by incorporating the detection of $J = 3$ into our state recovery sequence. To this end, efficient method to pump the $J = 3$ manifold must be developed. After pumping, the probabilities in $J = 3$ can be transferred to $J = 1$ via a Raman transition, as we have demonstrated in Ref. \cite{chou2020frequency} using a pair of frequency combs.

\bigskip
\noindent{\underline{Applicability of the state tracking and control protocol to other molecular ion species}}

To apply the state tracking and control protocol demonstrated here to another molecular ion species, efficient access of molecular state information is essential. Currently this is realized by transferring state information via the coupled harmonic motion between the ions. The closer the coupled motion is to the quantum mechanical ground state the better signal-to-noise ratio in molecular state detection. This results in a limitation on the mass-to-charge ratio ($m/z$) of the molecular ion for a given atomic ion, since in systems of large $m/z$ mismatches certain motional modes are not amenable to cooling to the ground state. However, recent experiments have demonstrated the transfer of excitations in modes not amenable to ground state cooling to those which are in systems of atomic ions with large $m/z$ mismatches \cite{hou2024indirect}. This represents a promising approach to overcoming the $m/z$ limitations in single molecule QLS experiments. 

“Signature” transitions with distinctive frequencies also improve signal-to-noise ratio of the detections and enable state-specific operations as well as polarization selectivity. While such transitions exist within rotational manifolds, transitions between rotational manifolds and even vibrational manifolds for a molecular species of interest could be considered for this purpose. The transition strengths and lifetimes of the involved states will need to accommodate the coherent operations on the molecule.

\end{document}